\documentclass[final,3p,times]{elsarticle}

\usepackage{amsmath}
\usepackage{mathrsfs}
\usepackage{amsfonts}
\usepackage{amssymb}
\usepackage{amsthm}
\usepackage{bm}
\usepackage{graphicx}
\usepackage{subfigure}
\usepackage{float}
\usepackage{epsfig}
\usepackage{color}
\usepackage{epstopdf}
\usepackage{pdfpages}
\usepackage{mathtools}
\usepackage{multirow,bigdelim}
\usepackage{scalerel}
\usepackage{xcolor}
\usepackage{blindtext}
\usepackage[linesnumbered,ruled,vlined]{algorithm2e}
\usepackage{romannum}
\usepackage{nomencl}
\usepackage{booktabs}
\usepackage{cuted}

\makenomenclature
\DeclarePairedDelimiter\abs{\lvert}{\rvert}%

\newcommand{\nosemic}{\renewcommand{\@endalgocfline}{\relax}}
\newcommand{\dosemic}{\renewcommand{\@endalgocfline}{\algocf@endline}}
\let\oldnl\nl
\newcommand{\nonl}{\renewcommand{\nl}{\let\nl\oldnl}}

\usepackage{etoolbox}
\makeatletter
    \patchcmd{\@author}{\global\let\@fnmark\@empty}{\global\let\@fnmark\@empty\global\let\@corref\@empty}{}{\@latex@error{Failed to patch \string\@author for \string\@corref reset}}
\makeatother

\journal{Applied Energy}

\begin{document}

\begin{frontmatter}



\title{Values of Coordinated Residential Space Heating in Demand Response Provision}


\author{Zihang Dong}
\author{Xi Zhang\corref{cor1}}
\cortext[cor1]{Corresponding author.}
\ead{x.zhang14@imperial.ac.uk}
\author{Goran Strbac}

\address{Department of Electrical and Electronic Engineering, Imperial College London, SW7 2AZ, UK}


\begin{abstract}
Demand side response from space heating in residential buildings can potentially provide huge amount of flexibility for the power system, particularly with deep electrification of the heat sector. In this context, this paper presents a novel distributed control strategy to coordinate space heating across numerous residential households with diversified thermal parameters. By employing an iterative algorithm under the game-theoretical framework, each household adjusts its own heating schedule through demand shift and thermal comfort compensation with the purpose of achieving individual cost savings, whereas the aggregate peak demand is effectively shaved on the system level. Additionally, an innovative thermal comfort model which considers both the temporal and spatial differences in customised thermal comfort requirement is proposed. Through a series of case studies, it is demonstrated that the proposed space heating coordination strategy can facilitate effective energy arbitrage for individual buildings, driving 13.96\% reduction in system operational cost and 28.22\% peak shaving. Moreover, the superiority of the proposed approach in thermal comfort maintenance is numerically analysed based on the proposed thermal comfort quantification model.

\end{abstract}




\end{frontmatter}


\renewcommand{\nomgroup}[1]{
	\item[\bfseries 
	\ifstrequal{#1}{A}{\emph{Indices and sets}}{%
	\ifstrequal{#1}{B}{\emph{Sets}}{%
	\ifstrequal{#1}{C}{\emph{Parameters}}{%
	\ifstrequal{#1}{D}{\emph{Variables}}{	            \ifstrequal{#1}{E}{\emph{Signals}}{
	\ifstrequal{#1}{F}{\emph{Functions}}{
	}}}}}}%
	]}

\newcommand{\nomunit}[1]{%
\renewcommand{\nomentryend}{\hspace*{\fill}#1}}

\nomenclature[A,01]{$j \in \mathscr{H}$}{Index and set of households.}
\nomenclature[A,02]{$t \in \mathscr{T}$}{Index and set of time steps.}




\nomenclature[C,01]{$\Delta t$}{Time discretization step. \nomunit{[h]}}
\nomenclature[C,03]{$\bar{P}_{j}$}{Rated heating power of household $j$. \nomunit{[kW]}}
\nomenclature[C,10]{$\overline{T}_{j,t}, \underline{T}_{j,t}$}{Preset upper/lower temperature bounds of household $j$ at time step $t$. \nomunit{[$^{\circ}\text{C}$]}}
\nomenclature[C,12]{$T^{ref}_{j,t}$}{Reference temperature of household $j$ at time step $t$. \nomunit{[$^{\circ}\text{C}$]}}

\nomenclature[C,12]{$T^{o}_{j,t}$}{Outdoor temperature of household $j$ at time step $t$. \nomunit{[$^{\circ}\text{C}$]}}


\nomenclature[C,13]{$D^{nh}_{t}$}{Non-heating demand at time step $t$. \nomunit{[MW]}}

\nomenclature[C,14]{$P^{res}_{t}$}{Renewable energy source power output at time step $t$. \nomunit{[MW]}}


\nomenclature[D,01]{$Q^{a}_{j,t}$}{Aggregate heat gains of building $j$ at time step $t$. \nomunit{[kW]}}
\nomenclature[D,0110]{$\epsilon_{j,t}$}{ON/OFF status of the heating device of household $j$ at time $t$. \nomunit{[p.u.]}}


\nomenclature[D,012]{$D^{h}_{t}$}{Heating demand at time step $t$. \nomunit{[MW]}}
\nomenclature[D,013]{$T_{j,t}$}{Temperature of household $j$ at time $t$. \nomunit{[$^{\circ}\text{C}$]}}
\nomenclature[D,02]{$p_{t}$}{Electricity price at time $t$. \nomunit{[$\text{\pounds}$/MWh]}}
\nomenclature[D,03]{$D_{t}$}{Aggregate electricity demand at time $t$. \nomunit{[MW]}}


\nomenclature[D,50]{$\bm{\sigma}_{j}$}{Matrix of demand shift operation. \nomunit{[p.u.]}}


\nomenclature[D,52]{$(t_{1}, t_{2})$}{Time pair for the demand shift (element column and row index in a matrix). \nomunit{[h]}}


\nomenclature[D,53]{$\bm{\theta}_{j}$}{Vector of additional heating operation. \nomunit{[p.u.]}}

\nomenclature[D,54]{$t_{0}$}{Time instance for the additional heating (element index in a vector). \nomunit{[h]}}

\nomenclature[D,542]{$\bm{\mathcal{C}}^{\sigma}_{j}$}{Matrix of electricity cost variation by demand shift. \nomunit{[$\text{\pounds}$]}}

\nomenclature[D,543]{$\bm{\mathcal{C}}^{\theta}_{j}$}{Vector of additional electricity cost due to thermal comfort compensation. \nomunit{[$\text{\pounds}$]}}

\nomenclature[D,55]{$\mathcal{E}_{j}$}{Thermal comfort index of household $j$. \nomunit{[$^{\circ}\text{C}$]}}





\nomenclature[E,03]{$\bm{\epsilon}_{j}$}{Heating profile $\epsilon_{j} = \{\epsilon_{j,1}, \cdots, \epsilon_{j,N_{T}}\}$ of household $j$ over time interval $\mathscr{T}$. \nomunit{[p.u.]}}
\nomenclature[E,03]{$\bm{T}_{j}$}{Temperature $T_{j} = \{T_{j,1}, \cdots, T_{j,N_{T}}\}$ of household $j$ over time interval $\mathscr{T}$. \nomunit{[$^{\circ}\text{C}$]}}

\nomenclature[F,10]{$f(\cdot)$}{Generation cost function.}
\nomenclature[F,101]{$g(\cdot)$}{Marginal cost function.}
\nomenclature[F,11]{$C_{T}(\cdot)$}{Total system cost function.}
\nomenclature[F,17]{$\Pi(\cdot)$}{Electricity marginal cost function.}

\printnomenclature

\section{Introduction}
\label{Int}

The electrification of residential space heating is widely accepted as an effective solution for the decarbonization of the future energy system, particularly in the scenario of high renewable energy sources (RES) penetration \cite{strbac2018analysis,heinen2018heat,wei2019electrification}.
Since heat demand is characterized by considerable volatility with substantial peak-to-valley gaps, once fully electrified, it will bring significant burden to power system expansion on both the generation and distribution sides. In this context, it is important to propose a smart control strategy to effectively utilize the flexibility in the heat sector such that electrification of the heat demand can operationally benefit the power system without incurring increased reinforcement burden.

Electrified space heating load is essentially a type of thermostatically controlled loads (TCLs). 
Considering the inherent inertia in a heat system, TCLs are typically hysterically controlled within pre-set temperature bounds. These temperature bounds are the key measure of flexibility that can be taken advantage of in the power system. Lots of literature has demonstrated the potential of using TCLs to collectively provide various ancillary services for the power system. 
In \cite{wei2018bi}, a bi-level scheduling approach for aggregating a large number of TCLs was presented,
while a two-stage optimization model of the distribution network considering the coordination of multiple TCL groups was proposed in \cite{wei2018coordination}.
A linear optimization model was established in \cite{trovato2016leaky} for the provision of frequency response 
through refrigerators, while demand side response through collective operation of diversified TCLs was proposed in \cite{trovato2017role}.
There are some other literature focused on controller design which enables TCLs to accurately deliver some functionalities.
For instance, centralised TCL controllers were constructed to provide operating reserve in \cite{lu2012design} and perform load tracking in \cite{hu2016load}, while distributed control approaches were applied in \cite{tindemans2015decentralized} for the controller design which enables each TCL to target a reference power profile. 




In the aforementioned works, TCL operation is restricted by fixed temperature settings, i.e., the upper and lower bounds do not change with time. However, residential space heating, an important type of TCL, is typically characterised by variable temperature settings. Specifically, when people are out of home, temperature restrictions can be relaxed since thermal comfort is not a concern during these periods; when people are awake at home, temperature restrictions will be tighter and may vary with time based on individual preference. Additionally, unlike other types of TCLs,
thermal comfort is particularly important for space heating. That is to say the delivery of any services through space heating should not substantively reduce the heating quality. These characteristics significantly increase the control complexity of space heating, considering the temporal and spatial diversity in temperature settings and time-dependent thermal comfort requirement across numerous households. Adequate works were focused on strategic scheduling of space heating with diversified objectives. Reference \cite{jin2017hierarchical} proposed a energy management method in an office building to reduce the daily heating cost, while \cite{yu2018energy} demonstrated a control approach to jointly optimize energy costs and indoor air quality of a building. Authors in \cite{li2020collaborative} analysed the benefits of thermal inertia in buildings based on an optimization-based heating model. Extensively, some works aim to manage space heating by considering more complicated building structures with multiple zones to improve the control accuracy \cite{blum2019practical,joe2019model,li2021study}. Moreover, model predictive control (MPC) strategies were used for space heating management, for example, to minimize building energy costs in \cite{bianchini2019integrated}, to reduce peak demand in \cite{ma2012demand}, and to enhance energy efficiency in \cite{serale2018model}, etc. However, the above works
were based on individual buildings, either considering single zone or multiple zones.

 
Regarding the control of multiple buildings, authors in \cite{zhang2013aggregated} developed an aggregation model aimed at managing a large population of air conditioners to provide frequency regulation, while \cite{adhikari2018algorithm} provides a control algorithm to optimally manage the aggregated demand of multiple HVAC units. Relevantly, the economic benefits of deploying heat storage in numerous buildings considering both the national and local constraints was addressed in \cite{dong2021evaluation}. Moreover, the alternating direction method of multipliers (ADMM) algorithm has been widely used in controlling multiple agents. For instance, the advantages of consensus-based ADMM for energy management problems were demonstrated in \cite{li2020new}, while a sharing ADMM algorithm was used to optimize the response of multiple TCLs in \cite{burger2017generation}. Although a lot of works have investigated the control approaches of demand response from multiple buildings, most of them cannot properly take into account the maintenance of thermal comfort. Specifically, they either assumes that thermal comfort is fully delivered if there is no temperature violation \cite{bunning2020experimental}, or considers the thermal comfort as a whole measurement covering the entire time horizon without differentiating the temporal difference, for instance, maintaining thermal comfort by minimizing the standard deviation of indoor temperature relative to the reference temperature \cite{calvino2010comparing,nguyen2014optimal,kou2020scalable,good2015optimization,ghilardi2021co}. 

In this context, this paper is dedicated to proposing a smart control approach to coordinate the space heating of a large population of residential buildings. By using this approach, individual households can effectively perform energy arbitrage for cost saving without essentially compromising heating quality while the power system can benefit from peak saving driven by demand side response of numerous households. An iteration-based algorithm is designed to sequentially adjust the binary heating schedule of individual households under a game-theoretic framework. By using this algorithm, the total generation cost converges to an equilibrium where no individual household can further drive cost reduction by unilaterally updating its own space heating schedule. Additionally, to consider the temporal difference in thermal comfort requirement, this paper introduces the innovative concept of comfort flexibility, which is fundamentally measured by the allowable temperature range of a particular time interval. 
Based on this concept, thermal comfort of low-flexibility periods is effectively maintained by compromising heating quality of high-flexibility periods, which is less concerned.

Compared to the existing literature, the key novelties and contributions of this paper are summarized as follows:
\begin{itemize}
    \item A novel distributed control approach is proposed to coordinate the binary space heating schedule of numerous residential buildings. Compared to most of the existing works where the electricity price is not influenced by demand response, this paper models the residential space heating as a cost-responsive load under a game-theoretic framework, such that individual households can directly interact with the generation side through marginal generation costs. Therefore, the proposed approach can effectively consider a wider system with a large population of participants, the collective demand response of which will fundamentally distort the electricity price.
    \item Based on the physical essence of heating arbitrage that increasing energy consumption beforehand to keep the thermal comfort maintained at a lower cost, we, for the first time, transform space heating into two independent processes, i.e., demand shift and thermal comfort compensation. Compared to most of the existing works where space heating is considered in an optimization-based model, the proposed approach transforms the heating optimisation problem to an iterative matrix calculation problem, thus significantly reducing the computational burden and increasing the calculation flexibility.
    \item An innovative thermal comfort quantification model is established to consider both the temporal and spatial differences in thermal comfort requirement across different households. Compared to the existing approaches where thermal comfort requirements over different periods are not differentiated, the proposed model prioritises the thermal comfort maintenance during comfort-inflexible periods by compromising the heating quality during comfort-flexible periods, which can intelligently adapts to individual settings of thermal comfort. 
\end{itemize}



The remainder of this paper is structured as follows: Section \ref{sec:iterativecontrolstrategy} introduces the basic models and assumptions, including the building thermal dynamic model and the simplification of the marginal electricity generation cost; Section \ref{sec:distributedpreheatingcoordinationscheme} proposes the algorithm of the coordinated space heating control strategy. Section \ref{IVNR} demonstrates the superior performance of the proposed control strategy through the numerical results of a number of case studies, while Section \ref{sec:conclusion} presents the conclusions and associated future work.

\section{Iterative control strategy}\label{sec:iterativecontrolstrategy}
\subsection{Thermal dynamic model of buildings}
\label{IIA}

\begin{figure}[H]
    \centering
    \includegraphics[width=0.3\linewidth]{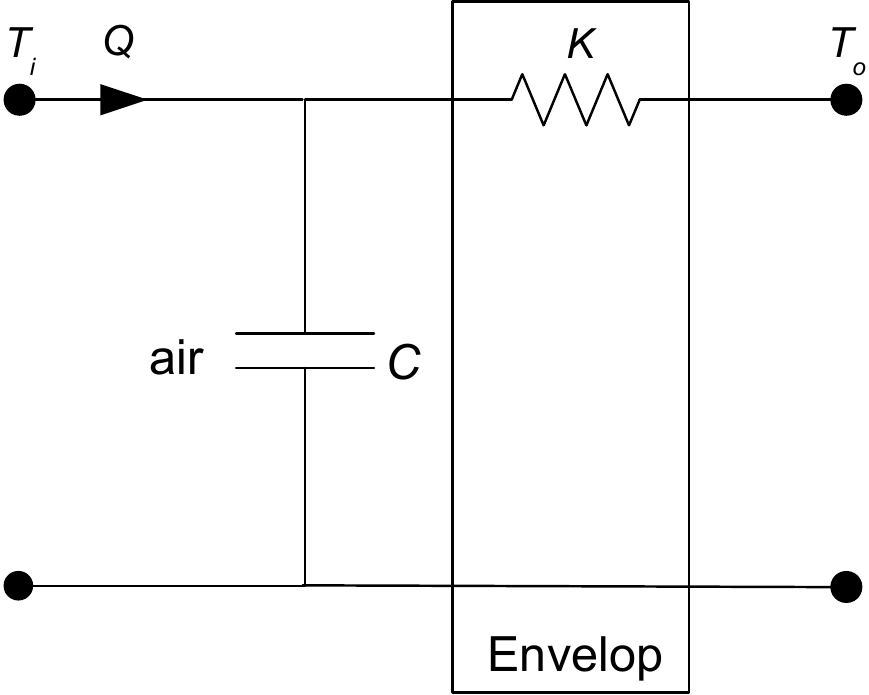}
    \caption{Equivalent electrical circuit of building thermal features}
    \label{fig:Therm-Circuit}
\end{figure}

The thermal functionality of a building can be compared to an electrical circuit. \figurename{~\ref{fig:Therm-Circuit}} illustrates a simplified first-order equivalent circuit to incorporate the key thermal features of residential buildings. Specifically, since the indoor air can store thermal energy, it is modelled as an equivalent thermal capacity $C^{eq}$. Additionally, heat loss constantly occurs through the building envelop, therefore, it can be regarded as an equivalent thermal conductance $K^{eq}$. 

Let us consider a set of households $\mathscr{H} = \{1, \cdots, H\}$ that can be represented by a first order circuit as in \figurename{~\ref{fig:Therm-Circuit}}. Then, the thermal dynamics of household $j$ can be governed by \eqref{eq1},
\begin{equation} \label{eq1}
    C^{eq}_{j} \frac{dT_{j}}{dt} = - (T_{j} - T_{j}^{o})K^{eq}_{j} + Q^{a}_{j}
\end{equation}
where $Q^{a}_{j}$ represents the aggregate thermal power injection in household $j$, including the power from heating appliance and passive heat gains. $T^{o}_{j}$ denotes the outdoor temperature, which fundamentally drives the heat loss.



Suppose the residential heating appliances are operated in binary model, i.e., work in either ``ON" or ``OFF" mode, equation \eqref{eq1} is discretized with sampling time interval $\Delta t$ and re-organized as \eqref{eq2},
\begin{equation} \label{eq2}
   T_{j,t+1}= \psi_{j}\cdot T_{j,t} + \gamma_{j} \cdot \epsilon_{j,t}+ \upsilon_{j,t}.
\end{equation}
where $t \in \mathscr{T} = \{1, \cdots , N_{T}\}$ represents the sampling time instants and $\epsilon_{j,t}$ is a binary variable representing the ``ON/OFF'' control action for the heating appliance. Specifically, $\epsilon_{j,t} = 1$ indicates the heating appliance is operating at its rated power, whereas $\epsilon_{j,t} = 0$ indicates the heating appliance is OFF. Note that $\psi_{j}$, $\gamma_{j}$ and $\upsilon_{j,t}$ are associated with thermal parameters of household $j$.

The indoor temperature $T_{j,t}$ are the state variables, and should be maintained within a preset range based on the personal preference of individual households, as shown in \eqref{const1},
\begin{equation}\label{const1}
     \underline{T}_{j,t} \leq T_{j,t} \leq \overline{T}_{j,t}.
\end{equation}

\subsection{Generation cost assumptions}
\label{EleDemandPrice}


Demand response from space heating can impact the system operational costs. Although the impact of a single participant is too minor to substantively change the marginal system operational cost, collective demand response performed by numerous consumers would fundamentally make a difference.

Basically, the marginal system operational cost is influenced by a series of factors, including electricity demand, RES availability, generator dynamic characteristics and transmission network topology, etc. To calculate the accurate marginal operational cost considering the impacts of all these factors, a detailed whole-system operation optimization model covering the generation side, transmission side and demand side should be established. Since the focus of this paper is to propose a novel control algorithm for collective residential heating, whereas the key driver of marginal generation costs is the net electricity demand\cite{kizilkale2012large,ma2016efficient}, for the convenience of demonstration without essentially impacting the conclusion, we skip the establishment of the detailed whole-system generation optimization problem and simplify the marginal cost function by only considering the impacts of net demand.

By convention, the electricity generation cost function is assumed to be quadratic in the form of \eqref{quad}. Then, the total generation cost can be defined as \eqref{ElectricityCost}, while the marginal generation cost is derived as \eqref{pricesig}.
\begin{equation}\label{quad}
    f(x) = \frac{1}{2}a\cdot x^2+b\cdot x,
\end{equation}

\begin{equation}\label{ElectricityCost}
    C_{T}(D) = \sum^{N_{T}}_{t=1} f(D_{t}) \Delta t
\end{equation}

\begin{equation}\label{pricesig}
    p_{t}= g(D_{t}) = \frac{\partial C_{T}}{\partial D_{t}}.
\end{equation} 



The net electricity demand $D$ is determined by non-heat electricity demand $D^{nh}$, heat-driven electricity demand $D^{h}$ and RES output $P^{res}$, as formulated in \eqref{eqDemandBal}.
Since numerous households are considered, it is assumed that any individual household $j \in \mathscr{H}$ cannot influence the marginal generation cost by unilaterally changing its demand, as shown in \eqref{eqPriAsump}.

\begin{equation}\label{netdemand}
    D^{n}_{t} = D^{nh}_t - P^{res}_{t}
\end{equation}

\begin{equation}\label{eqDemandBal}
     D_{t} = \sum^H_{j=1} \epsilon_{j,t}\cdot \bar{P}_{j} + D^{n}_{t}.
\end{equation}

\begin{equation}\label{eqPriAsump}
     g(D_{t}) = g(\sum_{k\in \mathscr{H}} \epsilon_{k,t}\cdot\bar{P}_{k} + D^{n}_{t}) \simeq g(\sum_{k\in \mathscr{H}\setminus\{j\}} \epsilon_{k,t}\cdot \bar{P}_{k} + D^{n}_{t})
\end{equation}

With the aim of delivering total generation cost savings, the proposed space heating coordination strategy is designed under the game-theoretic framework. Each household shifts its heat demand from periods with high marginal generation costs to periods with low marginal generation costs while considering thermal comfort maintenance. All participants interact with each other through the changes in marginal generation costs determined by collective heating strategies.

\section{Coordinated control strategy for collective residential space heating}\label{sec:distributedpreheatingcoordinationscheme}


Under the real-time electricity price scheme, the energy price can be influenced by building occupancy. Specifically, electricity price tends to rise when people are at home due to increased thermal comfort requirement and decrease when people are out of home. Therefore, people can shift space heating demand to the periods when they are out such that the remaining thermal energy can partially offset the heat demand when they get home, thus achieving energy arbitrage.
In this process, correct estimation of price signals is critical in guiding individual consumers to perform demand shifting which can effectively deliver energy savings. However, with a large population of households simultaneously adjusting their heating schedules, the original price signal would be severely distorted, thereby being invalid in guiding consumers to achieve energy arbitrage.
To handle this issue, we present a distributed control strategy to coordinate collective rescheduling of space heating across numerous households under a game-theoretical framework so that all players can be economically benefited by providing demand response.

\subsection{Demand shift}





\begin{figure}[H]
    \centering
    \includegraphics[width=0.6\textwidth]{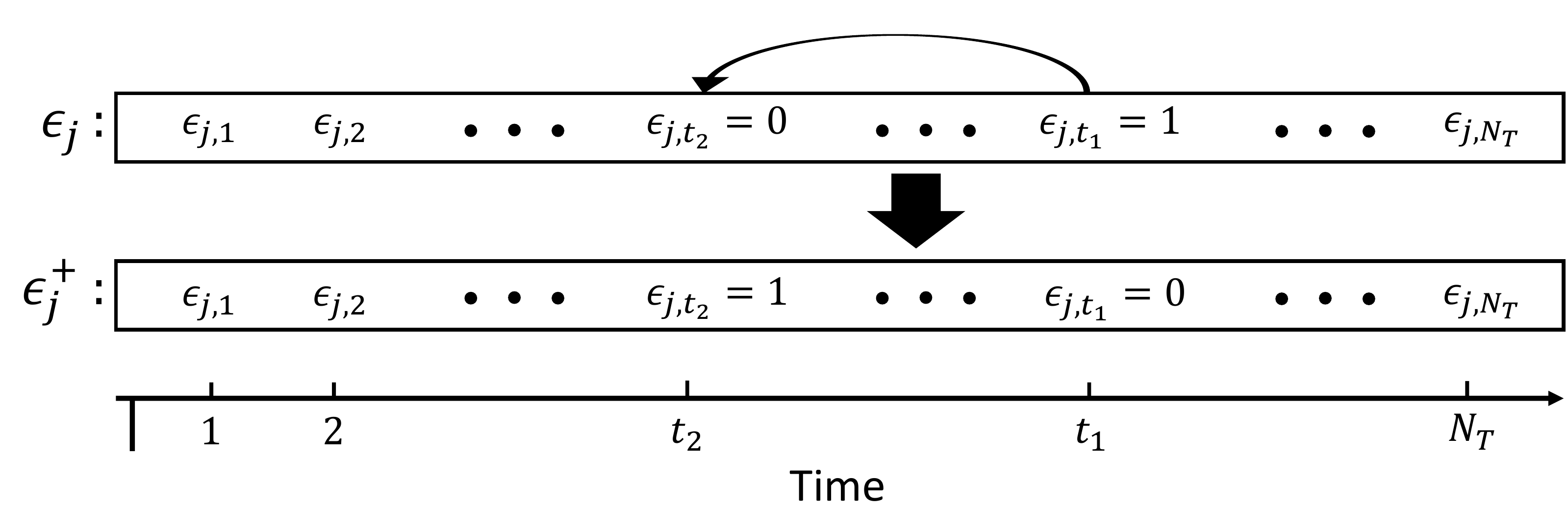}
    \vspace{-2mm}
    \caption{An example of demand shift operation from $t_{1}$ to $t_{2}$.}
    \label{fig:DemandShift}
\end{figure}

The proposed control strategy is composed of a sequence of ``demand shift" operations. Fig.~\ref{fig:DemandShift} demonstrates the process of demand shift from $t_{1}$ to $t_{2}$. Specifically, before demand shift, the heating schedule of household $j$ is denoted as $\epsilon_{j}$ where the heater is ``ON" at $t_{1}$ and ``OFF" at $t_{2}$. By performing demand shift, $\epsilon_{j}$ is changed to $\epsilon^{+}_{j}$ in which the heater status at $t_{1}$ and $t_{2}$ are exchanged. It should be stressed that demand shift can only be carried out from an ``ON" period to an ``OFF" period.
It is clear that demand shift cannot be implemented between any pair of time instants, i.e., $(t_{1}, t_{2})$. 
In this context, we introduce a ``feasibility factor", denoted as $\bm{\sigma}_{j}(t_{1},t_{2})$, to indicate whether demand shift from $t_{1}$ to $t_{2}$ is feasible. Specifically, $\bm{\sigma}_{j}(t_{1},t_{2}) = 1$ when heating demand can be shifted from $t_{1}$ to $t_{2}$ while $\bm{\sigma}_{j}(t_{1},t_{2}) = 0$ if demand cannot be shifted from $t_{1}$ to $t_{2}$.
For the convenience of narration, all demand shift feasibility factors are collected in a matrix $\bm{\sigma_j} \in \mathbb{R}^{N_{T}\times N_{T}}$ as expressed in \eqref{FFMat} for household $j$.

\begin{equation}\label{FFMat}
\bm{\sigma}_{j} = 
\begin{bmatrix}
0 & \bm{\sigma}_{j}(1,2) & \cdots & \bm{\sigma}_{j}(1,N_{T})\\
\bm{\sigma}_{j}(2,1) & 0 & \cdots & \bm{\sigma}_{j}(2,N_{T})\\
\vdots & \vdots & \ddots & \vdots\\
\bm{\sigma}_{j}(N_{T},1) & \bm{\sigma}_{j}(N_{T},2) & \cdots & 0
\end{bmatrix}
\end{equation}

Note that $\bm{\sigma_j}$ is determined by both physical restrictions of heating appliances and customized operational requirements. The following part of this subsection will elaborate how to determine $\bm{\sigma_j}$. To achieve so, three intermediate feasibility factor matrix considering different impacting factors are introduced: 


\subsubsection*{Operational status of heaters}

For a heating schedule $\bm{\epsilon}_{j}$, demand can be shifted from $t_{1}$ to $t_{2}$ only when $\epsilon_{j,t_{1}} = 1$ and $\epsilon_{j,t_{2}} = 0$. Thus, feasible demand shift operations are restricted by 
\begin{equation}\label{MaxPowerRestriction}
    \bm{\sigma}^{a}_{j}(t_{1},t_{2}) = \max \{0, \epsilon_{j,t_{1}} - \epsilon_{j,t_{2}}\}.
\end{equation}
where $\bm{\sigma^{a}_{j}}$ denotes the intermediate feasibility factor matrix considering the restriction of heater operational status.

\subsubsection*{Temperature requirement}

According to thermodynamic laws, demand shift from $t_{1}$ to $t_{2}$ will affect the temperature of all time instants after $\min\{t_{1}, t_{2}\}$. 
Therefore, it is essential to ensure that demand shift does not violate temperature constraints \eqref{const1}. To this end, let us denote a second intermediate feasibility factor matrix associated to temperature requirements by $\bm{\sigma^{b}_{j}}$. The element $\bm{\sigma}^{b}_{j}(t_1,t_2) = 1$ if shifting demand from $t_{1}$ to $t_{2}$ does not violate temperature constraints, otherwise, $\bm{\sigma}^{b}_{j}(t_1,t_2) = 0$.

To numerically calculate the values of $\bm{\sigma^{b}_{j}}$, we first reformulate the discretized thermal dynamic model \eqref{eq2} into a compact form \eqref{SystemEquation}:
\begin{equation} \label{SystemEquation}
    \bm{T}_{j} = \bm{\Psi}_{j} T_{j,0} + \bm{\Gamma}_{j}\bm{\epsilon}_{j} + \bm{\bm{\Upsilon}}_{j}
\end{equation}
where $\bm{T}_{j} = [T_{j,1}, T_{j,2}, \cdots, T_{j,N_{T}}]^{T}$ represents indoor temperature over the whole time horizon. Matrices $\bm{\Psi}_{j}$, $\bm{\Gamma}_{j}$, and vector $\bm{\Upsilon}_{j}$ are associated with the thermal parameters of household $j$, viz,
\begin{equation}
\bm{\Psi}_{j} =
    \begin{bmatrix}
    \psi_{j}\\
   \psi^{2}_{j} \\
   \vdots \\
    \psi^{N_{T}}_{j}
    \end{bmatrix},
    \bm{\Upsilon}_{j} = 
    \begin{bmatrix}
    \upsilon_{j,0} \\
    \psi_{j} \upsilon_{j,0} + \upsilon_{j,1} \\
    \vdots \\
    \sum^{N_{T}-1}_{\tau = 0} \psi^{N_{T}-\tau-1}_{j} \upsilon_{j,\tau}
    \end{bmatrix},\bm{\Gamma}_{j} =
    \begin{bmatrix}
    \gamma_{j} & 0 & \dots & 0 \\
    \psi_{j}\gamma_{j} & \gamma_{j} & \dots & 0 \\
    \vdots & \vdots & \ddots & \vdots \\
    \psi^{N_{T}-1}_{j}\gamma_{j} & \psi^{N_{T}-2}_{j}\gamma_{j} & \dots & \gamma_{j}
    \end{bmatrix}.
\end{equation}


In particular, the matrix $\bm{\Gamma}_{j}$ quantifies how the demand shift operation affects the temperature change. In other words, the temperature variation at any time instant $\tau \geq \min\{t_{1}, t_{2}\}$ due to the change of heater operation status at $t$ can be assessed via the element $\bm{\Gamma}_{j}(\tau,t)$. Specifically, switching off the heater at $t_{1}$ decreases the temperature $T_{j,\tau}$ by $\bm{\Gamma}_{j}(\tau,t_{1})$, whereas switching on the heater at $t_{2}$ increases the temperature $T_{j,\tau}$ by $\bm{\Gamma}_{j}(\tau,t_{2})$. As a result, the temperature at time $\tau$ will increase by $\bm{\Gamma}_{j}(\tau,t_{2}) - \bm{\Gamma}_{j}(\tau,t_{1})$ when the demand shift from $t_{1}$ to $t_{2}$ is feasible. Accordingly, elements in $\bm{\sigma^{b}_{j}}$ considering the restriction of temperature bounds are determined by \eqref{MaxSwap}: 
\begin{equation}\label{MaxSwap}
    \bm{\sigma}^{b}_{j}(t_{1},t_{2}) =
      \min \{1, \lfloor \hat{\sigma}^{b}_{j}(t_{1},t_{2}) \rfloor\}
\end{equation}
where the operator $\lfloor \cdot \rfloor$ denotes the nearest integer less than or equal to the variable, and $\hat{\sigma}^{b}_{j}(t_{1},t_{2})$ denotes the maximum demand that can be shifted which is determined by
\begin{equation}
    \hat{\sigma}^{b}_{j}(t_{1},t_{2}) = \underset{\tau \in \{\min\{t_{1},t_{2}\}+1,\cdots,N_{T}\}}{\text{min}}  \frac{\tilde{T}_{j,\tau,t_{1},t_{2}}}{\bm{\Gamma}_{j}(\tau,t_{2})-\bm{\Gamma}_{j}(\tau,t_{1})}
\end{equation}
where the maximum allowed temperature variation at time $\tau$ is given by \eqref{TempVar}:
\begin{equation}\label{TempVar}
    \tilde{T}_{j,\tau,t_{1},t_{2}}  = 
    \begin{cases}
    \underline{T}_{j,\tau}-T_{j,\tau} & \text{if}\,\, t_{1}<\tau \leq t_{2} \,\, \text{or}\,\,  t_{2}< t_{1} < \tau, \\
    \overline{T}_{j,\tau}-T_{j,\tau} & \text{if}\,\, t_{2}<\tau \leq t_{1} \,\, \text{or}\,\, t_{1}< t_{2} < \tau.
    \end{cases}
\end{equation}





\subsubsection*{Cost saving requirement}

From the perspective of individual households, the motivation of demand shift is to achieve cost savings, i.e., shifting demand from a high-price time instant $t_{1}$ to a low-price time instant $t_{2}$. However, demand shift can impact the energy price at related time instants, even reverse the original price order, i.e., $p_{t_{1}} > p_{t_{2}}$ and $p^{+}_{t_{1}} < p^{+}_{t_{2}}$. This can potentially lead to a situation where the demand shift fails in delivering cost savings as expected. This can further impact the convergence of the proposed iterative algorithm. To tackle this issue, we introduce a third intermediate matrix $\bm{\sigma}^{c}_{j}$ to ensure demand shift can always deliver cost savings. Numerically, the element at row $t_{1}$ and column $t_{2}$ is calculated as
\begin{equation}\label{AggregateDemandInequality}
    \bm{\sigma}^{c}_{j}(t_{1},t_{2}) = \min \{1, \lfloor \hat{\sigma}^{c}_{j}(t_{1},t_{2}) \rfloor \}
\end{equation}
where $\hat{\sigma}^{c}_{j}(t_{1},t_{2})$ is determined according to
\begin{equation}
    \hat{\sigma}^{c}_{j}(t_{1},t_{2}) = \frac{D_{t_{1}}-D_{t_{2}}}{2 \bar{P}_{j}} 
\end{equation}

The restriction in \eqref{AggregateDemandInequality} ensures that the updated aggregate demand satisfies
\begin{equation}\label{monotonedemand}
    D^{+}_{t_{1}} = D_{t_{1}} - \bar{P}_{j} \cdot \bm{\sigma}^{c}_{j}(t_{1},t_{2}) \geq D_{t_{2}} + \bar{P}_{j} \cdot \bm{\sigma}^{c}_{j}(t_{1},t_{2}) =  D^{+}_{t_{2}}.
\end{equation}
Additionally, \eqref{pricesig} implies that the marginal cost $p$ is monotonically increasing with respect to the aggregate electricity consumption $D$. Therefore, the inequality in \eqref{MonotonePrice} holds, indicating the updated marginal costs $p^{+}_{t_{2}}$ does not exceed $p^{+}_{t_{1}}$.
\begin{equation}\label{MonotonePrice}
    p^{+}_{t_{1}}=g(D^{+}_{t_{1}}) \geq g(D^{+}_{t_{2}})= p^{+}_{t_{2}}
\end{equation}
Thus, it has been demonstrated that the demand shift restricted by \eqref{AggregateDemandInequality} can preserve the order of marginal costs at $t_1$ and $t_2$, i.e. $p^{+}_{t_{2}}\leq p^{+}_{t_{1}}$, if $p_{t_{2}}< p_{t_{1}}$.

Up till now, considering all the restrictions above, i.e., \eqref{MaxPowerRestriction} for physical limits of heaters, and \eqref{MaxSwap} and \eqref{AggregateDemandInequality} for customized requirements, the ultimate matrix of demand shift feasibility factor is determined by 
\begin{equation}
    \bm{\sigma_{j}} = \bm{\sigma^{a}_{j}} \circ \bm{\sigma^{b}_{j}} \circ \bm{\sigma^{c}_{j}}.
\end{equation}
where the operator $\circ$ is the Hadamard product.


Then, the cost saving driven by demand shift can be expressed as \eqref{costvariation},
\begin{equation}\label{costvariation}
    \bm{\mathcal{C}}^{\sigma}_{j}(t_{1}, t_{2}) = \bar{P}_{j}\cdot (p_{t_{1}}-p_{t_{2}})\cdot \bm{\sigma}_{j}(t_{1},t_{2}) \cdot \Delta t
\end{equation}
which can be further expressed as the compact form of \eqref{costvariationmatrix}


\begin{equation}\label{costvariationmatrix}
    \bm{\mathcal{C}}^{\sigma}_{j} = \bar{P}_{j} \cdot \left(\bm{\Pi}\circ \bm{\sigma}_{j}\right) \cdot \Delta t,
\end{equation}
where $\bm{\Pi} \in \mathbb{R}^{N_{T}\times N_{T}}$ is calculated as \eqref{PDM}.

\begin{equation}\label{PDM}
\bm{\Pi} = 
\begin{bmatrix}
0 & p_{1} - p_{2} & \cdots & p_{1} - p_{N_{T}}\\
p_{2} - p_{1} & 0 & \cdots & p_{2} - p_{N_{T}}\\
\vdots & \vdots & \ddots & \vdots\\
p_{N_{T}} - p_{1} & p_{N_{T}} - p_{2} & \cdots & 0
\end{bmatrix}
\end{equation}

Given the matrix of cost savings $\bm{\mathcal{C}}^{\sigma}_{j}$ for demand shift between any pair of time instants $(t_{1},t_{2})$, we particularly select the pair that drives the most significant individual cost saving, i.e., \eqref{OptimalPairs}. This can facilitate the acceleration of the convergence of the iterative algorithm.
\begin{equation}\label{OptimalPairs}
    (t_{1}^{*}, t_{2}^{*})\in \underset{t_{1}, t_{2} \in \mathscr{T}}{\text{argmax}}\,\, \bm{\mathcal{C}}^{\sigma}_{j}(t_{1}, t_{2}).
\end{equation}

\subsection{Thermal comfort compensation}

\begin{figure}[H]
\vspace{-3mm}
\centering
\includegraphics[width=0.6\textwidth]{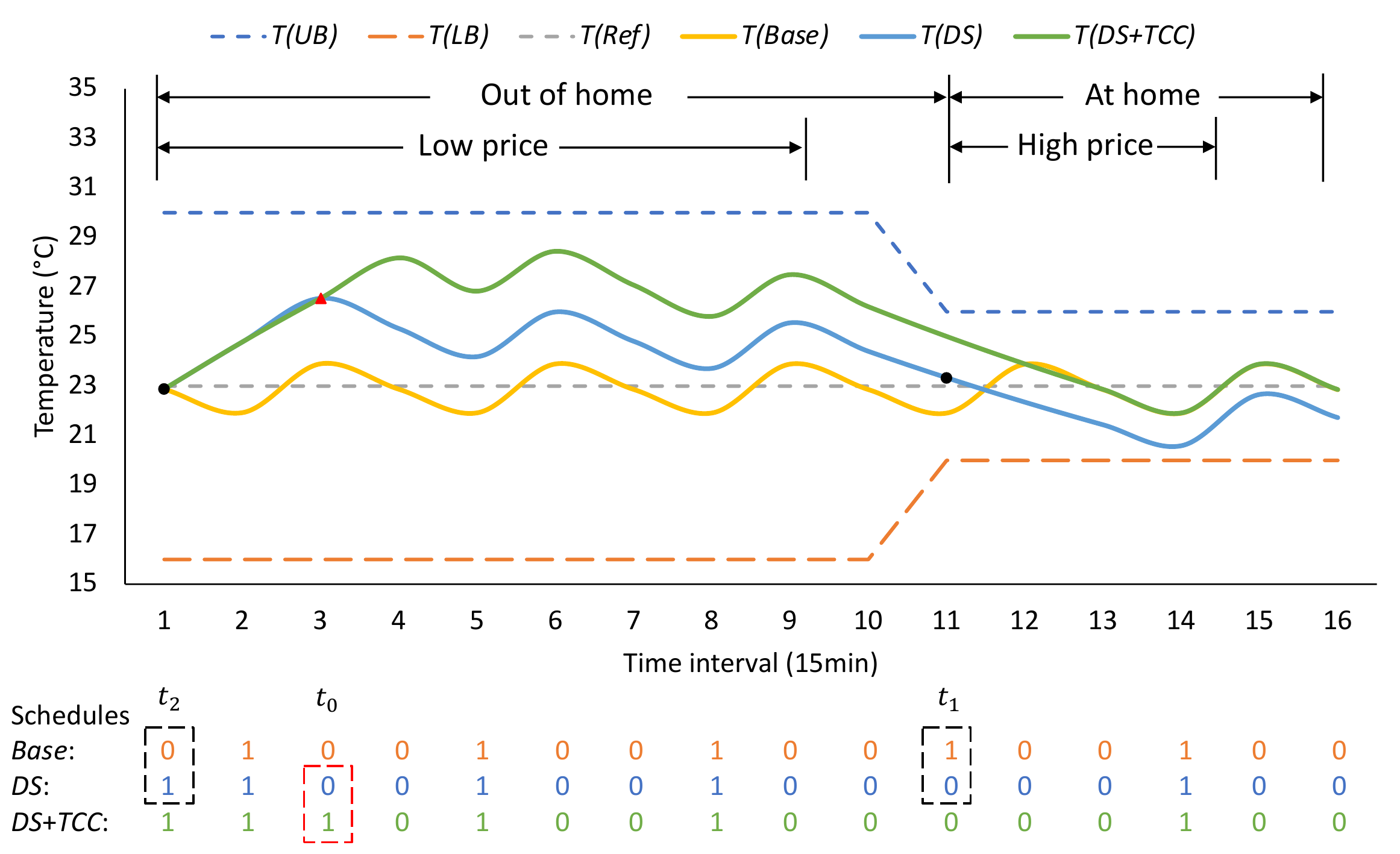}
\vspace{-3mm}
\caption{Illustration of thermal comfort compensation. $UB$: Upper Bound; $LB$: Lower Bound; $Ref$: Reference; $Base$: Baseline; $DS$: Demand Shift; $TCC$: Thermal Comfort Compensation.}
\label{fig:Process}
\end{figure}

The operation of demand shift can drive cost savings, however, it decreases the space heating quality during high-price periods. Although temperature settings are not violated, the overall thermal comfort is fundamentally compromised for high-price periods, which are typically correlated to demand-inflexible periods, e.g., morning and evening peaks. As shown in Fig.~\ref{fig:Process}, demand shift is performed from $t_{1}=11$ to $t_{2}=1$, which achieves cost savings. However, $T(DS)$ after $t=12$ is consistently lower than $T(Base)$, causing thermal comfort compromise when people are at home.
To compensate for the thermal comfort loss, we propose a strategy to overheat the room during low-price periods so that the inertial thermal energy can offset the later heat deficit during high-price periods where heating power is shifted out. Specifically in Fig.~\ref{fig:Process}, the heater is switched on at $t_{0}=3$ to narrow the gap between $T(DS)$ and $T(Base)$ after $t=12$, as illustrated by $T(DS+TCC)$. It is worth emphasizing that the temporal difference of thermal comfort requirement (reflected by the range between $T(UB)$ and $T(LB)$) during ``out of home" periods and ``at home periods" fundamentally enables the thermal comfort compensation process in Fig.~\ref{fig:Process}.

Before elaborating the detailed thermal comfort compensation strategy, let us first introduce the thermal comfort index under a given heating schedule $\epsilon_{j}$, as shown in \eqref{thermalcomfortmetric}, to quantify thermal comfort,
\begin{equation}\label{thermalcomfortmetric}
    \mathcal{E}_{j}(\epsilon_{j}) = - \frac{1}{N_{T}} \sum^{N_{T}}_{t = 1}  e_{j,t} \cdot \abs{ T_{j,t} - T^{ref}_{j,t} }
\end{equation}
where $e_{j,t}$ defines the thermal comfort flexibility and is expressed as
\begin{equation}\label{comfortflexibility}
    e_{j,t} = \frac{1}{\overline{T}_{j,t} - \underline{T}_{j,t}}.
\end{equation}

Note that $\mathcal{E}_{j}(\epsilon_{j})$ measures the overall temperature deviation of a given temperature profile with respect to the reference temperature considering the impacts of thermal comfort flexibility. 
Particularly, the temperature variation range, i.e., $\overline{T}_{j,t} - \underline{T}_{j,t}$, is used as an indicator of strictness of thermal comfort requirement. For instance, when out of home, people tend to set the range wide since indoor temperature is less concerned. When people are back home, thermal comfort is more demanding so the margin for temperature variation tends to be narrow. This index innovatively takes account of the temporal difference in thermal comfort requirement, which lays the foundation of smart energy arbitrage for individual households without essentially compromising space heating quality. 


Similar to the demand shift operation, it is necessary to determine the feasibility factors for thermal comfort compensation, denoted by vector $\bm{\theta}_{j}\in \mathbb{R}^{N_{T}}$, which is also affected by operational status of heaters and temperature requirement.

\subsubsection*{Operational status of heaters}

The thermal comfort compensation can only be carried out at a time instant where the heater is originally OFF. Therefore, the availability of thermal comfort compensation at $t_0$ for household $j$ is restricted by
\begin{equation}\label{TCCHeaterStat}
    \bm{\theta}^{a}_{j}(t_{0}) = 1 - \epsilon_{j,t_{0}}
\end{equation}
where $\bm{\theta}^{a}_{j}$ denotes the intermediate feasibility vector considering the restriction of heater operational status.

\subsubsection*{Temperature requirement}

Since switching on the heater at $t_{0}$ will increase the temperature for all the following periods, the thermal comfort compensation should be restricted by the temperature upper bound. To take this requirement into account, a second intermediate vector $\bm{\theta}^{b}_{j} \in \mathbb{R}^{N_{T}}$ is introduced. Specifically, $\bm{\theta}^{b}_{j}(t_{0}) = 1$ if the resulting temperature does not violate the bounds after thermal comfort compensation. Otherwise, $\bm{\theta}^{b}_{j}(t_{0}) = 0$. Mathematically, it is formulated by \eqref{TempInc},
\begin{equation}\label{TempInc}
    \bm{\theta}^{b}_{j}(t_{0}) = \min \{1, \lfloor \hat{\theta}^{b}_{j}(t_{0}) \rfloor\}
\end{equation}
where $\hat{\theta}^{b}_{j}(t_{0})$ quantifies the maximum power that can be increased at time $t_{0}$ considering the temperature requirement, viz.,
\begin{equation}
    \hat{\theta}^{b}_{j}(t_{0}) = \min_{\tau \in \{t_{0}+1,\cdots,N_{T}\}} \frac{\overline{T}_{j,\tau}-T_{j,\tau}}{\bm{\Gamma}_{j}(\tau,t_{0})}.
\end{equation}

Combining \eqref{TCCHeaterStat} and \eqref{TempInc}, the ultimate feasibility factors for thermal comfort compensation are determined by \eqref{UltTCCFF}, 
\begin{equation}\label{UltTCCFF}
    \bm{\theta}_{j} = \bm{\theta}^{a}_{j}\circ \bm{\theta}^{b}_{j}
\end{equation}
while the extra energy cost for thermal comfort compensation is expressed as
\begin{equation}
    \bm{\mathcal{C}}^{\theta}_{j} = \bar{P}_{j} \cdot  (\bm{p} \circ \bm{\theta}_{j}) \cdot \Delta t
\end{equation}
where $\bm{p} =  [p_{1}, p_{2}, \cdots, p_{N_{T}}]^{T}$ is the vector of electricity price and the element at $t_{0}$ is calculated as
\begin{equation}
\bm{\mathcal{C}}^{\theta}_{j}(t_{0}) = \bar{P}_{j}\cdot  p_{t_{0}} \cdot \bm{\theta}_{j}(t_{0}) \cdot \Delta t.
\end{equation}

Since there can be multiple time instants for feasible implementation of thermal comfort compensation, i.e., at time $t_0$ where $\bm{\mathcal{C}}^{\theta}_{j}(t_0)=1$, we specifically select $t^{*}_{0}$ which can compensate for the loss of thermal comfort in the most cost-effective way, viz., \begin{equation}\label{OptimalAddTime}
    t^{*}_{0}\in \underset{t_{0} \in \mathcal{T}:\, \bm{\theta}_{j}(t_{0})=1, \bm{\mathcal{C}}^{\theta}_{j}(t_{0}) < \bm{\mathcal{C}}^{\sigma}_{j}(t^{*}_{1},t^{*}_{2})}{\text{argmax}}\,\, \frac{\Delta \mathcal{E}_{j}(t_{0})}{ \bm{\mathcal{C}}^{\theta}_{j}(t_{0})}.
\end{equation}

Note that the constraints imposed in \eqref{OptimalAddTime} guarantee 1) the thermal comfort compensation is feasible, and 2) the additional cost incurred by thermal comfort compensation is lower than the cost saving by demand shift, thus achieving overall saving. Moreover, $\Delta \mathcal{E}_{j}(t_{0})$ is the increased thermal comfort due to the compensation at $t_{0}$ and it is calculated as \eqref{thermalcomfortvariation},
\begin{equation}\label{thermalcomfortvariation}
    \Delta \mathcal{E}_{j}(t_{0}) = \mathcal{E}_{j}(\epsilon^{+}_{j}) - \mathcal{E}_{j}(\epsilon_{j})  = \frac{1}{N_{T}} \sum^{N_{T}}_{t = t_{0}}  e_{j,t}\cdot  \Bigg( \abs{ T_{j,t} - T^{ref}_{j,t} }  - \abs{ T_{j,t} + \sum^{t-1}_{\tau = t_{0}} \bm{\Gamma}_{j}(\tau,t_{0}) - T^{ref}_{j,t}}\Bigg)
\end{equation}
where $\epsilon_{j}$ and $\epsilon^{+}_{j}$ represent the heating schedule before and after thermal comfort compensation at $t_{0}$.

\subsection{Summary of the algorithm}

Based on the operation of demand shift and thermal comfort compensation, the coordinated residential space heating strategy is summarised in four steps as follows:

\noindent \textit{Step 1}: Each household sets up a baseline heating schedule based on their preference and inform the central controller. The central controller calculates the electricity price and aggregate demand based on all the collected heating schedules, then send these information to the first household.


\noindent \textit{Step 2}: The household which received the information from the central controller updates its heating schedule by implementing demand shift and thermal comfort compensation, then informs the central controller of the updates.


\noindent \textit{Step 3}: The central controller revises the electricity price and aggregate demand, then communicates to the next household.


\noindent \textit{Step 4}: Repeat Step 2 and 3 until no household can further perform the demand shift, i.e., no household can further reduce the total system operational cost by unilaterally changing its heating schedule.



The space heating coordination process for numerous households is further demonstrated in the form of Algorithm \ref{IterativeAlgorithm}:

\begin{algorithm}
\DontPrintSemicolon 

\textbf{Initialization:}

\nonl  $\ell = 0$,\,\, $k = 0$,\,\, $conv = 0$, \,\, $\mathcal{E}^{(0)}_{j}= \mathcal{E}_{j}$

\While{$conv == 0$}
{
    $conv = 1$, $\ell = \ell + 1$
    
    \For{j=1:1:H}
    {
        $\epsilon^{(\ell)}_{j} = \epsilon^{(\ell-1)}_{j}$, $T^{(\ell)}_{j} = T^{(\ell-1)}_{j}$
        
        \textbf{Demand shift:}

        \textbf{Find} $(t^{*}_{1}, t^{*}_{2})$ 
        \textbf{do}:
        
        $conv = 0$
        
        $\epsilon^{(\ell)}_{j,t^{*}_{2}} = 1$, $\epsilon^{(\ell)}_{j,t^{*}_{1}} = 0$, and update $T^{(\ell)}_{j}$ 

            

            

            \textbf{Thermal comfort compensation:}
            
            \textbf{Find} $t^{*}_{0}$ \textbf{do:}
            
            
            $\epsilon^{(\ell)}_{j,t^{*}_{0}} = 1$ and update $T^{(\ell)}_{j}$ 

    }
    
}



    
    
    
    
    
    
    
    

\textbf{Result:} each household can achieve energy saving by demand shift and thermal comfort compensation

\caption{Coordinated space heating}
\label{IterativeAlgorithm}
\end{algorithm}

\section{Numerical Results}\label{IVNR}

This section is devoted to testing the proposed space heating coordination strategy in a series of case studies. The 24-hour horizon is discretized into 96 sampling time intervals, i.e., $\Delta t = 15$ min. The total number of households equipped with electric heaters is $2 \times 10^{5}$. All simulations were implemented using MATLAB R2019a, on a computer with 2-core 3.50GHz Intel(R) Xeon(R) E5-1650 processor and 32GB RAM.

\subsection{Parameters and assumptions}
\label{Para&Assumption}

\begin{figure}[H]
\vspace{-3mm}
\centering
\includegraphics[width=0.6\textwidth]{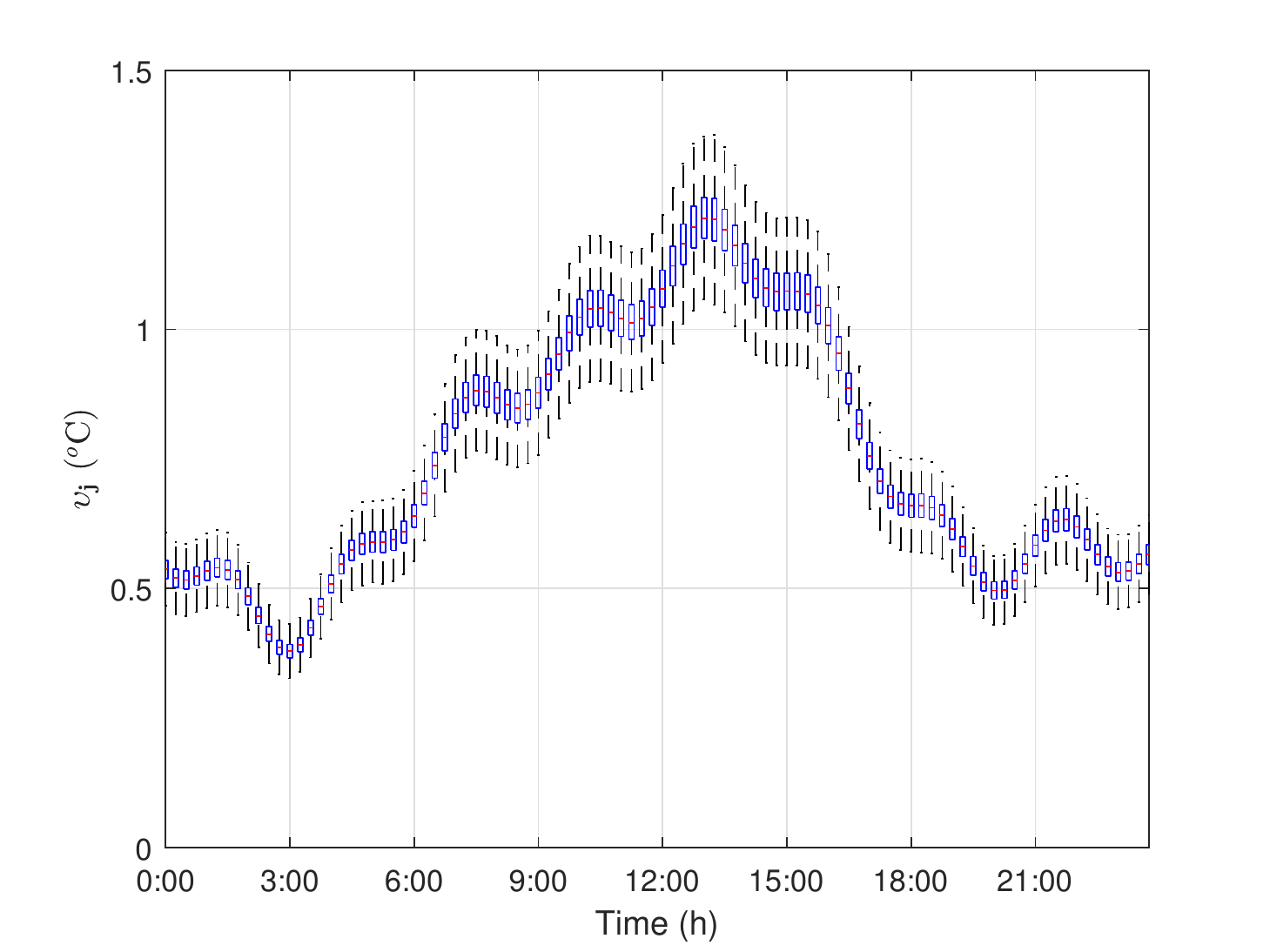}
\vspace{-3mm}
\caption{The box plot of thermal parameter $\upsilon_{j}$.}
\label{fig:upsilon}
\end{figure}

In the following case studies, the relevant thermal parameters $\psi_{j}$ and $\gamma_{j}$ in \eqref{eq2} are assumed to follow the Gaussian distribution (mean 0.94 and standard deviation 0.05 for $\psi_{j}$, mean 2 and standard deviation 0.1 for $\psi_{j}$). The time varying profiles of parameter $\upsilon_{j,t}$ over all households are statistically visualized by the box-plot in \figurename{~\ref{fig:upsilon}}. Particularly, the central red mark indicates the median, and the bottom and top edges of the box indicate the 25$^{\text{th}}$ and 75$^{\text{th}}$ percentiles, respectively. Moreover, the coefficients of generation cost function are assumed as $a = 0.03$ \pounds/MW$^2$h and $b = 12$ \pounds/MWh.

\begin{figure}[H]
\vspace{-3mm}
\centering
\includegraphics[width=0.6\textwidth]{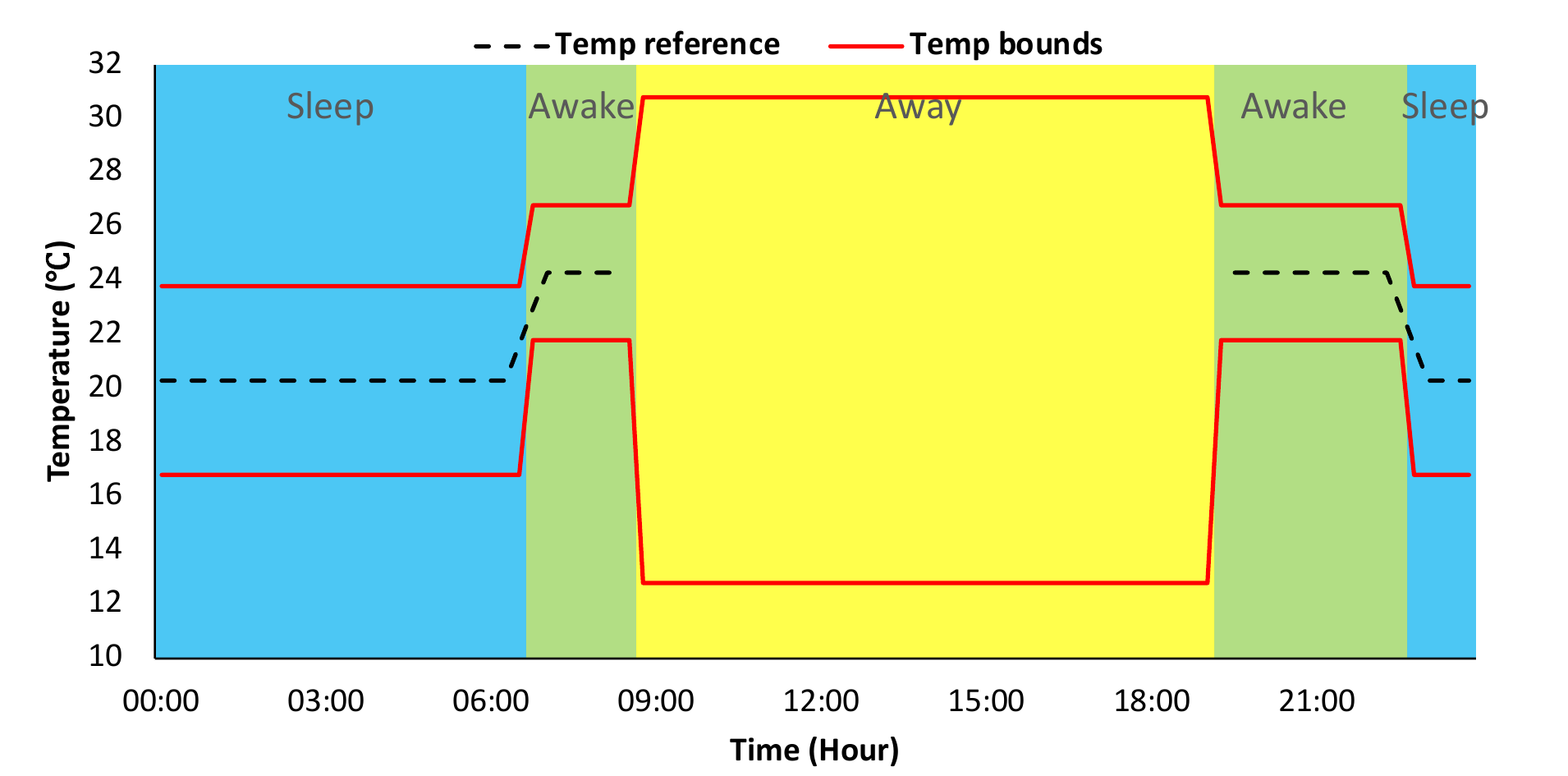}
\vspace{-3mm}
\caption{Mean values of temperature parameters.}
\label{fig:TemperatureParameters}
\end{figure}

As shown in \figurename{~\ref{fig:TemperatureParameters}}, three types of periods characterized by different thermal comfort flexibility including sleeping (blue), awake at home (green) and away from home (yellow) are differentiated. 
During the ``Awake" periods, the thermal comfort requirement is high thereby the comfort flexibility is low. Whereas during the ``Away" periods, thermal comfort is not a major concern so the comfort flexibility is rather high. Regarding the ``Sleep" periods, the thermal comfort requirement is in the between, indicating that people are less sensitive to the variation of temperature when sleeping. Note that the boundaries of three periods follow Gaussian distributions with the mean demonstrated in \figurename{~\ref{fig:TemperatureParameters}} and the standard deviation of 1 hour. The temperature bounds for these periods also follow the Gaussian distribution with the mean shown in \figurename{~\ref{fig:TemperatureParameters}} and the standard deviation of 1 $^\circ$C. Moreover, the reference temperature is determined as 
\begin{equation}\label{referencetemperature}
    T_{j,t}^{ref}=\frac{1}{2}\left(\underline{T}_{j,t} + \overline{T}_{j,t}\right), \quad t \in \mathcal{T}^{in}_{j}
\end{equation} 
where $\mathcal{T}^{in}_{j} \subseteq \mathcal{T}_{j}$ denotes the time when people are at home. $T_{j,t}^{ref}$ reflects the desired thermal comfort which is inactive when no occupants are at home.

It should be stressed that the case studies aim at providing a demonstration of the superiority of the proposed distributed preheating strategy. All input parameters can be changed based on realistic data and individuals' preference in real applications without loss of effectiveness of the methodology.

\subsection{Case studies: performance of the algorithm}\label{casestudy1}
To evaluate the performance of the proposed space heating coordination strategy, the following cases are investigated:

\textit{Baseline case}: Space heating is controlled to track the reference temperature $T^{ref}_{j}$.


\textit{Case 1}: Space heating in different households is not coordinated. Individual households independently minimize their electricity bills based on the baseline price.

\textit{Case 2}: Space heating in all households is coordinated by using the proposed control strategy.



\begin{figure}[H]
\vspace{-3mm}
\centering
\includegraphics[width=0.6\textwidth]{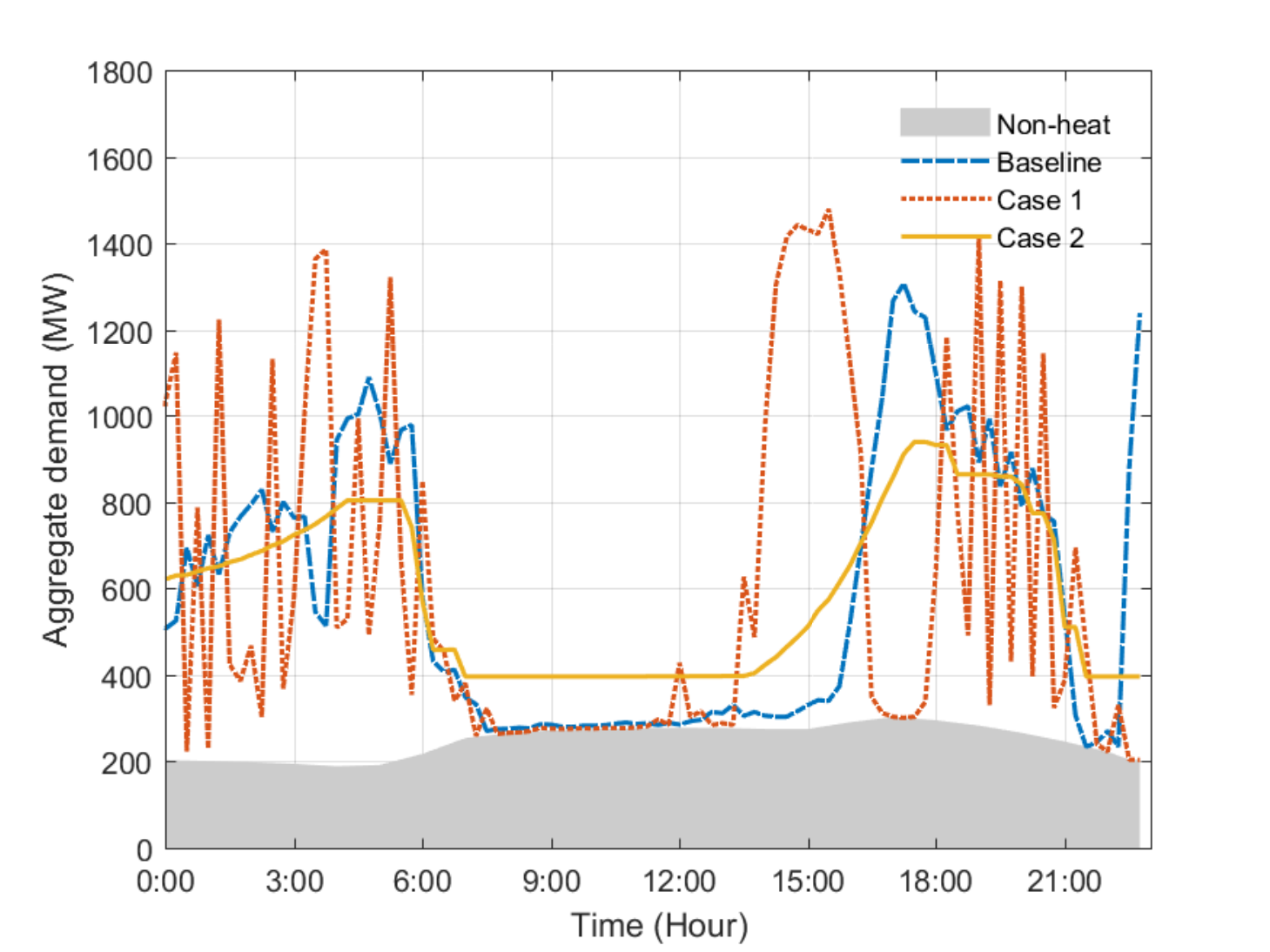}
\vspace{-3mm}
\caption{Aggregate demand profiles using different algorithms.}
\label{fig:AggregateDemand}
\end{figure}

The aggregate demand profiles in different cases are illustrated in \figurename{~\ref{fig:AggregateDemand}}. It is observed that the baseline demand profile, represented by the blue curve, is characterized by a morning peak and an evening peak. A deep valley between the two peaks corresponds to the period when people are out for work and shows that there is almost no heating appliance switched on. In Case 1 where the heating-driven energy arbitrage is performed without coordination, the original peaks are shaved, but new peaks are formed right before. Additionally, strong oscillations can be observed in the aggregate demand caused by simultaneous heating, which pose huge threats to power system operation. In contrast, the aggregate demand profile in Case 2 is significantly improved in the favour of system operation. Specifically, the original peaks are effectively shaved, i.e., 26.22\% for the morning peak and 28.22\% for the evening peak. Moreover, the volatility in the original demand profile is substantially reduced. Through this case study, the superior performance of the proposed space heating coordination strategy in improving the power system operating condition has been demonstrated. 

\begin{figure}[H]
\vspace{-3mm}
\centering
\includegraphics[width=0.6\textwidth]{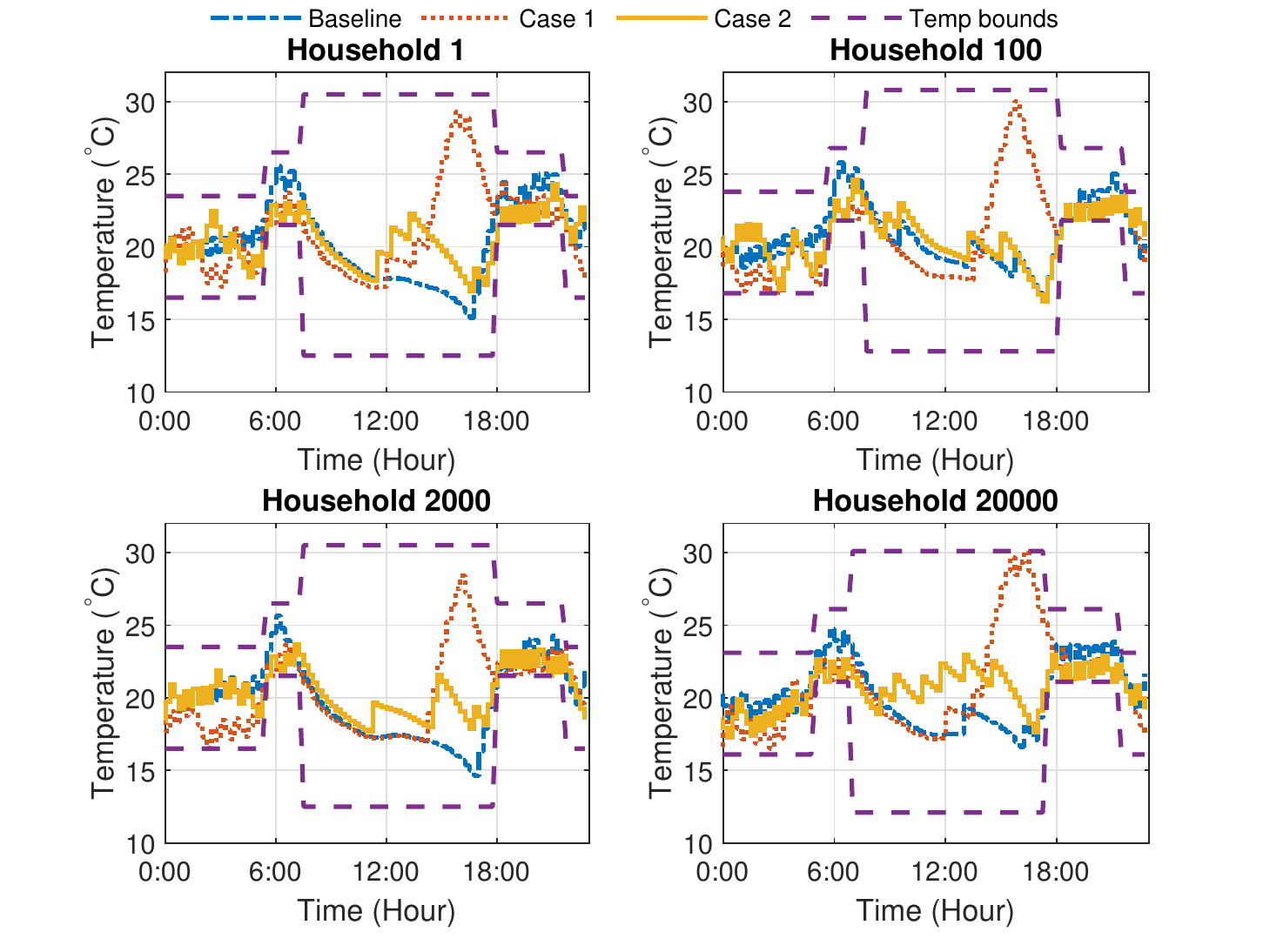}
\vspace{-3mm}
\caption{Indoor temperature of households under binary heating mode.}
\label{fig:TemperatureBinary}
\end{figure}

\begin{figure}[H]
\vspace{-3mm}
\centering
\includegraphics[width=0.6\textwidth]{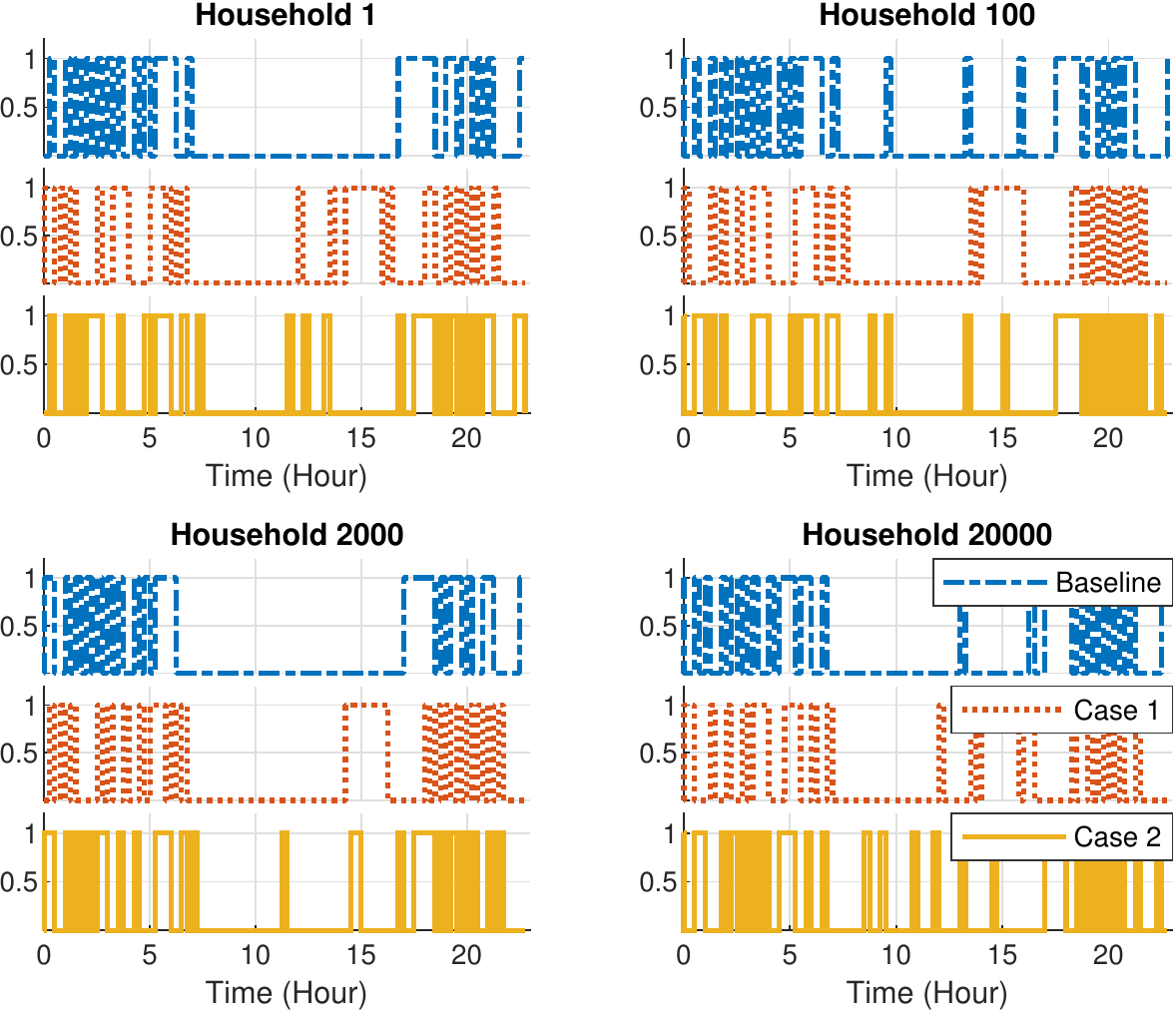}
\vspace{-3mm}
\caption{Heating schedule of households under binary heating mode.}
\label{fig:ScheduleBinary}
\end{figure}

Detailed information on the household level is further given in \figurename{~\ref{fig:TemperatureBinary}} and  \figurename{~\ref{fig:ScheduleBinary}}, which demonstrate the indoor temperature and the corresponding heating schedule respectively in 4 randomly selected households. For the baseline temperature with blue color in \figurename{~\ref{fig:TemperatureBinary}}, indoor temperature tracks the reference temperature when people are at home and drifts downward when people are out. Red curves highlight Case 1 where space heating is optimized individually without coordination while yellow curves depict Case 2 where the proposed coordination strategy is used. These temperature profiles are corresponding to the heating patterns in \figurename{~\ref{fig:ScheduleBinary}}. Note that 1 on the vertical axis indicates ``ON" and 0 indicates ``OFF". It can be seen that different households tend to follow a similar heating schedule when coordination is absent. This is because individual heating optimizations are uniformly guided by the original electricity price profile, which can be distorted when a large number of households change their heating schedules. In contrast, when the coordinated control strategy is applied as in Case 2, the space heating patterns in different households are diversified so that the peaks are shaved and the valleys are filled.

\begin{figure}[H]
\vspace{-3mm}
\centering
\includegraphics[width=0.6\textwidth]{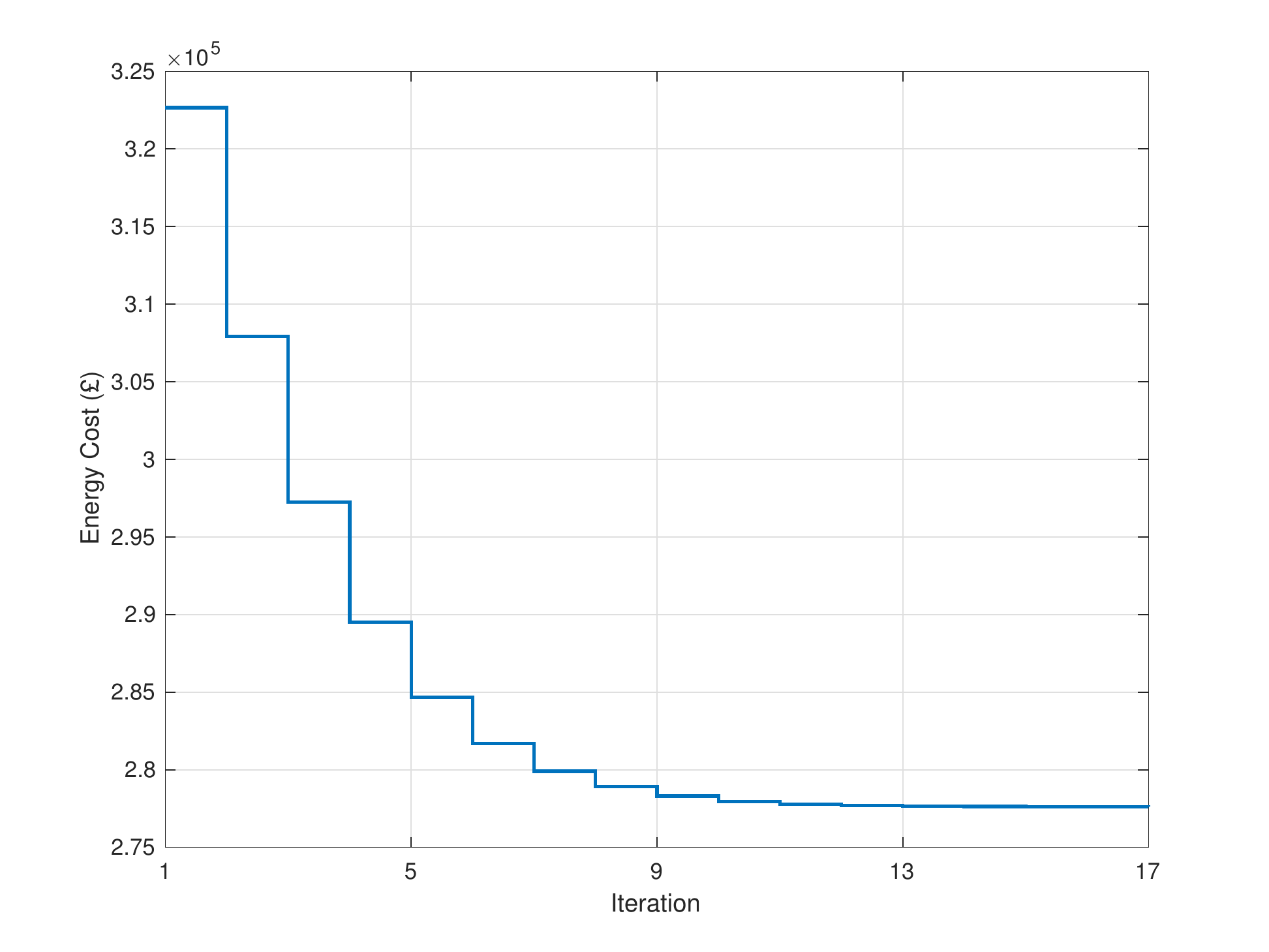}
\vspace{-3mm}
\caption{Total generation costs against the number of iteration.}
\label{fig:TotalCost}
\end{figure}

The convergence performance of Case 2 is illustrated in \figurename{~\ref{fig:TotalCost}}. As can be seen, the generation cost converges after 20 iterations and takes around 1.2 hours. Note that the number of iteration corresponds to the value of $\ell$ in Algorithm \ref{IterativeAlgorithm} which represents an adjustment on the heating schedule over all households. The total energy cost is reduced by 13.96\% compared to the baseline case and by 25.35\% compared to Case 1.

\subsection{Case studies: thermal comfort compensation}


This case study is dedicated to demonstrating the performance of the proposed approach regarding thermal comfort compensation. For this purpose, three cases are compared with the baseline case in Section \ref{casestudy1}:


\textit{Case A}: Only demand shift is conducted and there is no thermal comfort compensation.

\textit{Case B}: Temporal difference in thermal comfort requirement is neglected, i.e., comfort flexibility $e_{j,t}$ in the calculation of thermal comfort index is constant over the day as in \eqref{constantflexibility}
\begin{equation}\label{constantflexibility}
    e_{j,t} = \text{constant}.
\end{equation}

\textit{Case C}: The temporal difference in thermal comfort requirement is taken into account, i.e., the comfort flexibility $e_{j,t}$ is defined as \eqref{comfortflexibility}.

\begin{figure}[H]
\vspace{-3mm}
\centering
\includegraphics[width=0.6\textwidth]{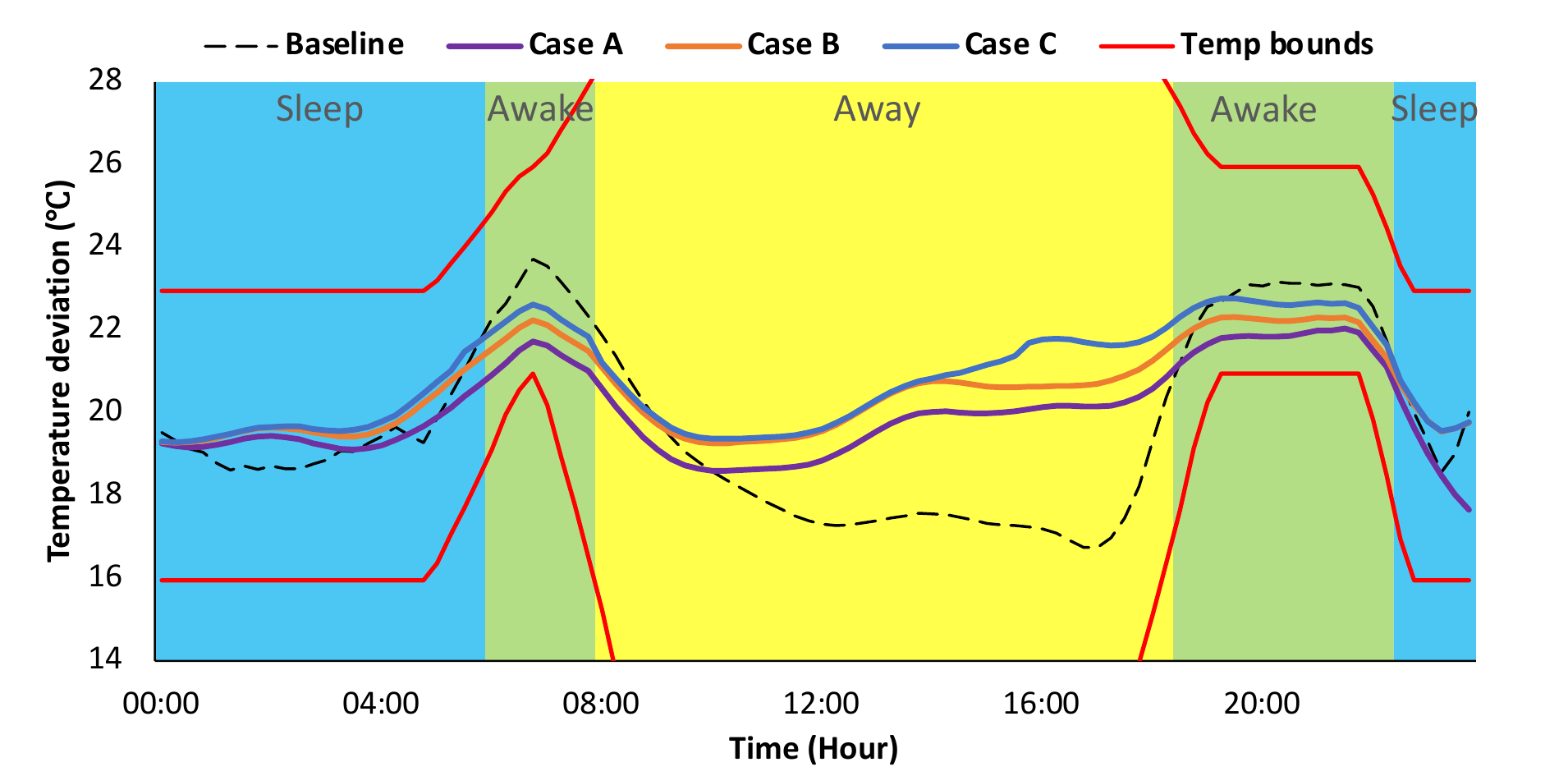}
\vspace{-3mm}
\caption{Average temperature profiles of the three cases.}
\label{fig:TemperatureAve}
\end{figure}

Since the demand response of individual households to the price signal can be case-specific, we calculate the average temperature conditions of 1000 households to statistically demonstrate the difference in thermal comfort of different cases, as in \figurename{~\ref{fig:TemperatureAve}}.
For Case A where only demand shift is performed, the temperature of low comfort flexibility periods decreases compared to the baseline case because the heating demand is shifted to the periods with higher comfort flexibility and lower electricity price. As a result, a large gap can be observed between the temperature of Case A and the baseline case during ``Awake" periods.
After considering the thermal comfort compensation, both Case B and Case C can narrow the gap, but Case C shows better performance in compensating the thermal comfort loss incurred by demand shift.  


\section{Conclusion}\label{sec:conclusion}

In this paper, a novel space heating coordination strategy is introduced for a large population of households equipped with electrified heating appliances operating in binary mode. Specifically, on the demand side, individual households update their heating schedules iteratively through demand shift and thermal comfort compensation to achieve cost savings without essentially satisfying heating quality, while on the system side, peak shaving is effectively realised through the coordination of space heating in numerous households, thus driving significant operational benefits. 

Through a series of case studies, the superiority of the space heating coordination strategy in peak shaving and cost saving is demonstrated. Based on the numerical results, the total generation cost is reduced by $13.96\%$ and the peak demand is reduced by $28.22\%$ compared to the baseline case where individual households do not perform energy arbitrage. Meanwhile, if the coordination strategy is absent where individual households perform energy arbitrage independently, $15.26\%$ cost increase is incurred compared to the baseline case while the peak demand is $13.11\%$ higher. Furthermore, the advantages of the proposed thermal comfort compensation mechanism that considers both the temporal and spatial comfort flexibility in households are illustrated. 

The proposed space heating coordination is targeted at day-ahead heating scheduling which shows the potential capability of demand side response from residential space heating loads. In the future, it will be valuable to investigate the real-time control of residential space heating considering various uncertainties.

\appendix
\section{Proof of convergence}
\begin{proof}[\unskip\nopunct]

This appendix is dedicated to proving that there is no household can further reduce the total system operational cost after all households reach the converged solution. 

Let us first consider the demand shift of an arbitrary iteration $\ell$ from $t^{*}_{1}$ to $t^{*}_{2}$, the heating status before and after this operation are 
\begin{equation}
\begin{aligned}
    \epsilon^{(\ell-1)}_{j,t^{*}_{1}} = 1 \to \epsilon^{(\ell)}_{j,t^{*}_{1}} = 0,& \quad
    \epsilon^{(\ell-1)}_{j,t^{*}_{2}} = 0 \to \epsilon^{(\ell)}_{j,t^{*}_{2}} = 1, \\
    \epsilon^{(\ell)}_{j,t} = \epsilon^{(\ell-1)}_{j,t},& \quad \forall t \in \mathcal{T} \setminus \{t^{*}_{1}, t^{*}_{2}\}.
\end{aligned}
\end{equation}
Accordingly, the aggregate demands formulated in \eqref{eqDemandBal} at these two instants fulfil the following equations
\begin{equation}
\begin{aligned}
    & D^{(\ell)}_{t^{*}_{1}} + D^{(\ell-1)}_{t^{*}_{1}} = 2 D^{(\ell-1)}_{t^{*}_{1}} - \bar{P}_{j}, \quad
    D^{(\ell)}_{t^{*}_{1}} - D^{(\ell-1)}_{t^{*}_{1}} = - \bar{P}_{j} \\
    & D^{(\ell)}_{t^{*}_{2}} + D^{(\ell-1)}_{t^{*}_{2}} = 2 D^{(\ell-1)}_{t^{*}_{2}} + \bar{P}_{j}, \quad
    D^{(\ell)}_{t^{*}_{2}} - D^{(\ell-1)}_{t^{*}_{2}} = \bar{P}_{j}
\end{aligned}
\end{equation}
Then, the variation of total system operational cost after performing the demand shift of can be explicitly derived as \eqref{totalcostvariation}
\begin{equation}\label{totalcostvariation}
\begin{aligned}
    &C_{T}(D^{(\ell)}) - C_{T}(D^{(\ell-1)}) \\
    =& \sum^{T}_{t = 1} \Bigg[  \frac{1}{2} a \left( D^{(\ell)}_{t}  \right)^{2} - \frac{1}{2} a \left( D^{(\ell-1)}_{t}  \right)^{2} + b  \left( D^{(\ell)}_{t}  \right) - b  \left( D^{(\ell-1)}_{t}  \right) \Bigg]\\
    =& \frac{1}{2} a \left[  \left( D^{(\ell)}_{t^{*}_{1}}  \right)^{2} - \left( D^{(\ell-1)}_{t^{*}_{1}}  \right)^{2} \right] + \frac{1}{2} a \left[ \left( D^{(\ell)}_{t^{*}_{2}}  \right)^{2} -  \left( D^{(\ell-1)}_{t^{*}_{2}}  \right)^{2} \right]  + b \left[ \left( D^{(\ell)}_{t^{*}_{1}} \right) -  \left( D^{(\ell-1)}_{t^{*}_{1}}  \right)  \right] + b \left[ \left( D^{(\ell)}_{t^{*}_{2}} \right) -  \left( D^{(\ell-1)}_{t^{*}_{2}}  \right)  \right]\\
    =& \frac{1}{2} a \left( D^{(\ell)}_{t^{*}_{1}} + D^{(\ell-1)}_{t^{*}_{1}}  \right)\left( D^{(\ell)}_{t^{*}_{1}} - D^{(\ell-1)}_{t^{*}_{1}} \right) + \frac{1}{2} a \left( D^{(\ell)}_{t^{*}_{2}} + D^{(\ell-1)}_{t^{*}_{2}}  \right)\left( D^{(\ell)}_{t^{*}_{2}} - D^{(\ell-1)}_{t^{*}_{2}} \right)+ b \left[ \left( D^{(\ell)}_{t^{*}_{1}} \right) -  \left( D^{(\ell-1)}_{t^{*}_{1}}  \right)  \right] + b \left[ \left( D^{(\ell)}_{t^{*}_{2}} \right) -  \left( D^{(\ell-1)}_{t^{*}_{2}}  \right)  \right]\\
    =& \frac{1}{2} a \left(2 D^{(\ell-1)}_{t^{*}_{1}} - \bar{P}_{j} \right) \left( - \bar{P}_{j} \right) + \frac{1}{2} a \left(2 D^{(\ell-1)}_{t^{*}_{2}} + \bar{P}_{j} \right) \left( \bar{P}_{j} \right) + b \left( -\bar{P}_{j} \right) + b\left( \bar{P}_{j} \right) \\
    =& \frac{1}{2} a \left( \bar{P}_{j} \right) \left[ \left(2 D^{(\ell-1)}_{t^{*}_{2}} + \bar{P}_{j}\right) - \left(2 D^{(\ell-1)}_{t^{*}_{1}}- \bar{P}_{j} \right) \right] \\
    = & \frac{1}{2} a \left( \bar{P}_{j} \right) \left[ 2\left( D^{(\ell-1)}_{t^{*}_{2}} - D^{(\ell-1)}_{t^{*}_{1}} \right) + 2\bar{P}_{j}  \right]\\
    \leq & \frac{1}{2} a \left( \bar{P}_{j} \right) \left[ 2\left( -2 \bar{P}_{j} \right) + 2\bar{P}_{j}  \right]\\
    = & -a \left(\bar{P}_{j}\right)^{2} \leq 0
\end{aligned}
\end{equation}
where the first inequality is implied from the cost saving requirement for aggregate demand in \eqref{monotonedemand}. As a result, equation \eqref{totalcostvariation} yields a strict reduction of total system operational cost by performing a demand shift, i.e.,
\begin{equation}
    C_{T}(D^{(\ell)}) < C_{T}(D^{(\ell-1)}).
\end{equation}
Together with the fact the total cost is bounded below, the iterative demand shift operation ensures the convergence of the total operational cost.

Next, we are able to show that the converged solution can make sure that no household can achieve further total cost reduction by unilaterally changing its heating profile. To verify this conclusion, it is necessary prove the following equation:
\begin{equation}\label{equilibriumcondition}
\begin{aligned}
    \sum_{t\in\mathscr{T}} f \left(D^{*}_{t} \right) \cdot \Delta t = 
 &\min_{\epsilon}\,\,\sum_{t\in\mathscr{T}} f \left(D_{t} \right) \cdot \Delta t\\
 \text{s.t.}&\,\, D_{t} = \sum_{i \in \mathscr{H}} \left( \epsilon_{i,t}\cdot \bar{P}_{i,t} + D^{nh}_{i,t} \right) - P^{res}_{t} \\
 & \,\, \sum_{t \in \mathscr{T}} \epsilon_{i,t} = \sum_{t \in \mathscr{T}} \epsilon^{*}_{i,t},\; \; \forall i \in \mathscr{H}\\
 &\,\,\epsilon_{i} \in \mathscr{E}_{i},\; \forall i \in \mathscr{H}
 \end{aligned}
\end{equation}
where $\mathscr{E}_{i}$ is the collection of heating schedules satisfying temperature and heating conditions
\begin{equation}\label{FeasiblePowerSchedule}
    \mathscr{E}_{j} = \{\epsilon_{j} \,|\, \underline{T}_{j,t+1} \leq \psi_{j}\cdot T_{j,t} + \gamma_{j} \cdot \epsilon_{j,t}+ \upsilon_{j,t} \leq \overline{T}_{j,t+1}, 
    \epsilon_{j,t} \in \{0, 1\}, \forall t \in \mathscr{T}  \}
\end{equation}

Then, let us prove the satisfaction of \eqref{equilibriumcondition} by contradiction. Assume that there exists a household $j \in \mathscr{H}$ which can further reduce the total cost by updating its converged heating schedule $\epsilon^{*}_{j}$, while all the other households remain their converged schedules, viz.
\begin{equation}\label{newschedule}
    \epsilon_{i} = \begin{cases}
    \epsilon^{**}_{j} & \text{if}\,\, i = j\\
    \epsilon^{*}_{i} & \text{otherwise}
    \end{cases}
\end{equation}
where $\epsilon^{**}_{j}$ is an updated schedule for household $j$ which could lead to a lower total cost than $\epsilon^{*}_{j}$ or in another words equation \eqref{equilibriumcondition} does not hold. Following our demand shift operation, the heating schedule $\epsilon^{**}_{j}$ can be expressed as
\begin{equation}
    \epsilon^{**}_{j} = \epsilon^{*}_{j} + \sum^{S}_{s = 1} \sigma_{j}(t^{(s)}_{1}, t^{(s)}_{2})
\end{equation}
where $S$ is the number of demand shift changing from $\epsilon^{*}_{j}$ to $\epsilon^{**}_{j}$, and each $\epsilon^{(s)}$ corresponds to a demand shift from $t^{(s)}_{1}$ to $t^{(s)}_{2}$. Then, given the cost variation of a single demand shift operation in \eqref{totalcostvariation}, the heating schedule adjustment by household $j$ from $\epsilon^{*}_{j}$ to $\epsilon^{**}_{j}$ would result in a cost reduction calculated as
\begin{equation}\label{individualcostvariation}.
\begin{aligned}
    C_{T}(D^{**}) - C_{T}(D^{*}) = \frac{1}{2} a \cdot \sum^{S}_{s = 1} & \Bigg[ \bigg[ \sum_{i \in \mathscr{H}\setminus \{j\}} 2\epsilon^{(s)}_{i,t^{(s)}_{1}}\cdot\bar{P}_{i} + 2\left(D^{nh}_{t^{(s)}_{1}} - P^{res}_{t^{(s)}_{1}} \right) + \left(\epsilon^{(s)}_{j,t^{(s)}_{1}}+\epsilon^{(s-1)}_{j,t^{(s)}_{1}}\right)\cdot \bar{P}_{j}\bigg] \cdot \bigg[\left(\epsilon^{(s)}_{j,t^{(s)}_{1}}-\epsilon^{(s-1)}_{j,t^{(s)}_{1}}\right)\cdot \bar{P}_{j} \bigg]  \Bigg]\\
     +& \Bigg[ \bigg[ \sum_{i \in \mathscr{H}\setminus \{j\}} 2\epsilon^{(s)}_{i,t^{(s)}_{2}}\cdot\bar{P}_{i} + 2\left(D^{nh}_{t^{(s)}_{2}} - P^{res}_{t^{(s)}_{2}} \right) + \left(\epsilon^{(s)}_{j,t^{(s)}_{2}}+\epsilon^{(s-1)}_{j,t^{(s)}_{2}}\right)\cdot \bar{P}_{j}\bigg] \cdot \bigg[\left(\epsilon^{(s)}_{j,t^{(s)}_{2}}-\epsilon^{(s-1)}_{j,t^{(s)}_{2}}\right)\cdot \bar{P}_{j} \bigg]  \Bigg]
\end{aligned}
\end{equation}
where $D^{**}$ denotes the aggregate demand associated with the heating schedule $\epsilon^{**}_{j}$.

Based on the assumption by contradiction that $\epsilon^{**}_{j}$ leads to a lower total cost than $\epsilon^{*}_{j}$, there must exist at least one $s \in \{1,\cdots,S\}$ such that $C_{T}(D^{(s)}) < C_{T}(D^{*})$ where $D^{(s)}$ is the aggregate demand profile corresponding the heating schedule $\epsilon^{(s)}_{j}$. This indicates that there exists positive demand shift action and the iterative algorithm is still convenient after $\epsilon^{*}_{j}$ has been reached (the termination condition is not satisfied). However, this contradicts to the fact that $\epsilon^{*}$ is the converged result of the heating profile for all households.

\end{proof}





\section*{Acknowledgments}
This research is supported by the research project: Active Building Centre (ABC, Grant EP/V012053/1).

\bibliographystyle{ieeetr}
\bibliography{reference.bib}

\begin{thebibliography}{10}

\bibitem{strbac2018analysis}
G.~Strbac, D.~Pudjianto, R.~Sansom, P.~Djapic, H.~Ameli, N.~Shah, N.~Brandon,
  A.~Hawkes, and M.~Qadrdan, ``Analysis of alternative uk heat decerbonisation
  pathways,'' {\em Imperial College London: London, UK}, 2018.

\bibitem{heinen2018heat}
S.~Heinen, P.~Mancarella, C.~O'Dwyer, and M.~O'Malley, ``Heat electrification:
  The latest research in europe,'' {\em IEEE Power and Energy Magazine},
  vol.~16, no.~4, pp.~69--78, 2018.

\bibitem{wei2019electrification}
M.~Wei, C.~A. McMillan, {\em et~al.}, ``Electrification of industry: potential,
  challenges and outlook,'' {\em Current Sustainable/Renewable Energy Reports},
  vol.~6, no.~4, pp.~140--148, 2019.

\bibitem{wei2018bi}
C.~Wei, J.~Xu, S.~Liao, Y.~Sun, Y.~Jiang, D.~Ke, Z.~Zhang, and J.~Wang, ``A
  bi-level scheduling model for virtual power plants with aggregated
  thermostatically controlled loads and renewable energy,'' {\em Applied
  energy}, vol.~224, pp.~659--670, 2018.

\bibitem{wei2018coordination}
C.~Wei, J.~Xu, S.~Liao, Y.~Sun, Y.~Jiang, and Z.~Zhang, ``Coordination
  optimization of multiple thermostatically controlled load groups in
  distribution network with renewable energy,'' {\em Applied Energy}, vol.~231,
  pp.~456--467, 2018.

\bibitem{trovato2016leaky}
V.~Trovato, S.~H. Tindemans, and G.~Strbac, ``Leaky storage model for optimal
  multi-service allocation of thermostatic loads,'' {\em IET Generation,
  Transmission \& Distribution}, vol.~10, no.~3, pp.~585--593, 2016.

\bibitem{trovato2017role}
V.~Trovato, F.~Teng, and G.~Strbac, ``Role and benefits of flexible
  thermostatically controlled loads in future low-carbon systems,'' {\em IEEE
  Transactions on Smart Grid}, vol.~9, no.~5, pp.~5067--5079, 2017.

\bibitem{lu2012design}
N.~Lu and Y.~Zhang, ``Design considerations of a centralized load controller
  using thermostatically controlled appliances for continuous regulation
  reserves,'' {\em IEEE Transactions on Smart Grid}, vol.~4, no.~2,
  pp.~914--921, 2012.

\bibitem{hu2016load}
J.~Hu, J.~Cao, M.~Z. Chen, J.~Yu, J.~Yao, S.~Yang, and T.~Yong, ``Load
  following of multiple heterogeneous tcl aggregators by centralized control,''
  {\em IEEE Transactions on Power Systems}, vol.~32, no.~4, pp.~3157--3167,
  2016.

\bibitem{tindemans2015decentralized}
S.~H. Tindemans, V.~Trovato, and G.~Strbac, ``Decentralized control of
  thermostatic loads for flexible demand response,'' {\em IEEE Transactions on
  Control Systems Technology}, vol.~23, no.~5, pp.~1685--1700, 2015.

\bibitem{jin2017hierarchical}
X.~Jin, J.~Wu, Y.~Mu, M.~Wang, X.~Xu, and H.~Jia, ``Hierarchical microgrid
  energy management in an office building,'' {\em Applied energy}, vol.~208,
  pp.~480--494, 2017.

\bibitem{yu2018energy}
L.~Yu, D.~Xie, C.~Huang, T.~Jiang, and Y.~Zou, ``Energy optimization of hvac
  systems in commercial buildings considering indoor air quality management,''
  {\em IEEE Transactions on Smart Grid}, 2018.

\bibitem{li2020collaborative}
X.~Li, W.~Li, R.~Zhang, T.~Jiang, H.~Chen, and G.~Li, ``Collaborative
  scheduling and flexibility assessment of integrated electricity and district
  heating systems utilizing thermal inertia of district heating network and
  aggregated buildings,'' {\em Applied Energy}, vol.~258, p.~114021, 2020.

\bibitem{blum2019practical}
D.~Blum, K.~Arendt, L.~Rivalin, M.~Piette, M.~Wetter, and C.~Veje, ``Practical
  factors of envelope model setup and their effects on the performance of model
  predictive control for building heating, ventilating, and air conditioning
  systems,'' {\em Applied Energy}, vol.~236, pp.~410--425, 2019.

\bibitem{joe2019model}
J.~Joe and P.~Karava, ``A model predictive control strategy to optimize the
  performance of radiant floor heating and cooling systems in office
  buildings,'' {\em Applied Energy}, vol.~245, pp.~65--77, 2019.

\bibitem{li2021study}
Z.~Li and J.~Zhang, ``Study on the distributed model predictive control for
  multi-zone buildings in personalized heating,'' {\em Energy and Buildings},
  vol.~231, p.~110627, 2021.

\bibitem{bianchini2019integrated}
G.~Bianchini, M.~Casini, D.~Pepe, A.~Vicino, and G.~G. Zanvettor, ``An
  integrated model predictive control approach for optimal hvac and energy
  storage operation in large-scale buildings,'' {\em Applied energy}, vol.~240,
  pp.~327--340, 2019.

\bibitem{ma2012demand}
J.~Ma, J.~Qin, T.~Salsbury, and P.~Xu, ``Demand reduction in building energy
  systems based on economic model predictive control,'' {\em Chemical
  Engineering Science}, vol.~67, no.~1, pp.~92--100, 2012.

\bibitem{serale2018model}
G.~Serale, M.~Fiorentini, A.~Capozzoli, D.~Bernardini, and A.~Bemporad, ``Model
  predictive control ({MPC}) for enhancing building and hvac system energy
  efficiency: Problem formulation, applications and opportunities,'' {\em
  Energies}, vol.~11, no.~3, p.~631, 2018.

\bibitem{zhang2013aggregated}
W.~Zhang, J.~Lian, C.-Y. Chang, and K.~Kalsi, ``Aggregated modeling and control
  of air conditioning loads for demand response,'' {\em IEEE transactions on
  power systems}, vol.~28, no.~4, pp.~4655--4664, 2013.

\bibitem{adhikari2018algorithm}
R.~Adhikari, M.~Pipattanasomporn, and S.~Rahman, ``An algorithm for optimal
  management of aggregated hvac power demand using smart thermostats,'' {\em
  Applied Energy}, vol.~217, pp.~166--177, 2018.

\bibitem{dong2021evaluation}
Z.~Dong, X.~Zhang, and G.~Strbac, ``Evaluation of benefits through coordinated
  control of numerous thermal energy storage in highly electrified heat
  systems,'' {\em Energy}, p.~121600, 2021.

\bibitem{li2020new}
W.~Li, Y.~Liu, H.~Liang, and Y.~Shen, ``A new distributed energy management
  strategy for smart grid with stochastic wind power,'' {\em IEEE Transactions
  on Industrial Electronics}, vol.~68, no.~2, pp.~1311--1321, 2020.

\bibitem{burger2017generation}
E.~M. Burger and S.~J. Moura, ``Generation following with thermostatically
  controlled loads via alternating direction method of multipliers sharing
  algorithm,'' {\em Electric Power Systems Research}, vol.~146, pp.~141--160,
  2017.

\bibitem{bunning2020experimental}
F.~B{\"u}nning, B.~Huber, P.~Heer, A.~Aboudonia, and J.~Lygeros, ``Experimental
  demonstration of data predictive control for energy optimization and thermal
  comfort in buildings,'' {\em Energy and Buildings}, vol.~211, p.~109792,
  2020.

\bibitem{calvino2010comparing}
F.~Calvino, M.~La~Gennusa, M.~Morale, G.~Rizzo, and G.~Scaccianoce, ``Comparing
  different control strategies for indoor thermal comfort aimed at the
  evaluation of the energy cost of quality of building,'' {\em Applied Thermal
  Engineering}, vol.~30, no.~16, pp.~2386--2395, 2010.

\bibitem{nguyen2014optimal}
D.~T. Nguyen and L.~B. Le, ``Optimal bidding strategy for microgrids
  considering renewable energy and building thermal dynamics,'' {\em IEEE
  Transactions on Smart Grid}, vol.~5, no.~4, pp.~1608--1620, 2014.

\bibitem{kou2020scalable}
X.~Kou, F.~Li, J.~Dong, M.~Starke, J.~Munk, Y.~Xue, M.~Olama, and H.~Zandi, ``A
  scalable and distributed algorithm for managing residential demand response
  programs using alternating direction method of multipliers {(ADMM)},'' {\em
  IEEE Transactions on Smart Grid}, vol.~11, no.~6, pp.~4871--4882, 2020.

\bibitem{good2015optimization}
N.~Good, E.~Karangelos, A.~Navarro-Espinosa, and P.~Mancarella, ``Optimization
  under uncertainty of thermal storage-based flexible demand response with
  quantification of residential users’ discomfort,'' {\em IEEE Transactions
  on Smart Grid}, vol.~6, no.~5, pp.~2333--2342, 2015.

\bibitem{ghilardi2021co}
L.~M.~P. Ghilardi, A.~F. Castelli, L.~Moretti, M.~Morini, and E.~Martelli,
  ``Co-optimization of multi-energy system operation, district heating/cooling
  network and thermal comfort management for buildings,'' {\em Applied Energy},
  vol.~302, p.~117480, 2021.

\bibitem{kizilkale2012large}
A.~C. Kizilkale, S.~Mannor, and P.~E. Caines, ``Large scale real-time bidding
  in the smart grid: A mean field framework,'' in {\em 2012 IEEE 51st IEEE
  Conference on Decision and Control (CDC)}, pp.~3680--3687, IEEE, 2012.

\bibitem{ma2016efficient}
Z.~Ma, S.~Zou, L.~Ran, X.~Shi, and I.~A. Hiskens, ``Efficient decentralized
  coordination of large-scale plug-in electric vehicle charging,'' {\em
  Automatica}, vol.~69, pp.~35--47, 2016.

\end{thebibliography}
\biboptions{sort&compress}




\end{document}